\documentclass[]{aastex701}
\hypersetup{linkcolor=red,citecolor=blue,filecolor=cyan,urlcolor=magenta}
\usepackage{amsmath}

\def\micron{\hbox{\,$\mu$m}}
\def\etal.{\hbox{\textit{et al.}}}
\def\eg{\hbox{\textit{e.g.}}}
\def\ie{\hbox{\textit{i.e.}}}
\def\degr{\hbox{$^\circ$}}

\def\gk{2015~GK$_{56}$}
\def\data{{\mathcal D}}

\newcommand{\au}{au}   % Astronomical unit - if we want to change 
\newcommand{\eqq}[1]{Equation~(\ref{#1})}
\newcommand{\likeli}{\mathcal{L}}
\newcommand{\shape}{\mathcal{S}}

\shorttitle{Size distribution of cold classical TNOs}
\submitjournal{PSJ}
\begin{document}

\title{The size and mass distribution of cold classical TNOs for $5<H<13$}

%% \author[orcid=0000-0002-9072-1121,gname=Gregory,sname=Schwarz]{Greg Schwarz}
%%
%% \author
%% \affiliation
%% \email
%%
%% \author[orcid=0000-0002-9072-1121,gname=Gregory,sname=Schwarz]{Greg Schwarz}
%%
\author[orcid=0000-0002-8613-8259, gname=Gary,sname=Bernstein]{Gary M. Bernstein}
\affiliation{Department of Physics and Astronomy, University of Pennsylvania, Philadelphia, PA 19104, USA}
\email[show]{garyb@upenn.edu}

\author[orcid=0000-0001-6680-6558, gname=Wesley, sname=Fraser]{Wesley C. Fraser}
\affiliation{National Research Council of Canada, Herzberg Astronomy and Astrophysics Research Centre, 5071 W. Saanich Rd. Victoria, BC, V9E
2E7, Canada}
\affiliation{Department of Physics and Astronomy, University of Victoria, Elliott Building, 3800 Finnerty Road, Victoria, BC V8P 5C2, Canada}
\email{Wesley.Fraser@nrc-cnrc.gc.ca}

\author[orcid=0000-0002-0760-1584, gname=Marielle, sname=Eduardo]{Marielle R. Eduardo}
\affiliation{Department of Physics and Astronomy, University of Victoria, 3800 Finnerty Road, Victoria, BC V8P 5C2, Canada}
\email{meduardo@uvic.ca
}
\author[orcid=0000-0002-8296-6540, gname=William, sname=Grundy]{William M. Grundy}
\affiliation{Department of Astronomy and Planetary Science, Northern Arizona University, Flagstaff, AZ 86011, USA}
% \affiliation{Lowell Observatory, 1400 W Mars Hill Rd., Flagstaff, AZ, 86001, USA}
\email{will.grundy@nau.edu}

\author[orcid=0000-0002-6875-1543, gname=Bryan, sname=Hilbert]{Bryan Hilbert}
\affiliation{Space Telescope Science Institute, 3700 San Martin Drive, Baltimore, MD 21218, USA}
\email{hilbert@stsci.edu}

\author[orcid=0000-0002-1139-4880, gname=Matthew, sname=Holman]{Matthew J. Holman}
\affiliation{Center for Astrophysics | Harvard \& Smithsonian, 60 Garden Street, Cambridge, MA 02138, USA}
\email{mholman@cfa.harvard.edu}

\author[orcid=0000-0001-5932-9570,gname=Anastasia, sname=Morgan]{Anastasia N. Morgan}
\affiliation{Department of Astronomy and Planetary Science, Northern Arizona University, Flagstaff, AZ 86011, USA}
\email{anm744@nau.edu}

\author[orcid=0000-0003-4827-5049, gname=Kevin, sname=Napier]{Kevin J. Napier}
\affiliation{Department of Physics, University of Michigan, Ann Arbor, MI 48109, USA}
\affiliation{Center for Astrophysics | Harvard \& Smithsonian, 60 Garden Street, Cambridge, MA 02138, USA}
\email{kevin.napier@cfa.harvard.edu}

\author[orcid=0000-0003-2434-5225,gname=John, sname=Stansberry]{John A. Stansberry}
\affiliation{Space Telescope Science Institute, 3700 San Martin Drive, Baltimore, MD 21218, USA}
\email{jstans@stsci.edu}

\author[orcid=0000-0003-4580-3790, gname=David, sname=Trilling]{David E. Trilling}
\affiliation{Department of Astronomy and Planetary Science, Northern Arizona University, Flagstaff, AZ 86011, USA}
\email{david.trilling@nau.edu}

%% \correspondingauthor command.
%% The command appends "Corresponding Author: " to the argument it appears at
%% the bottom left of the first page like the output from \email. 

\begin{abstract} The cold classical trans-Neptunian objects (CCs) are the only observable \textit{in situ} population of planetesimal remnants believed to have escaped collisional grinding. 
Recent JWST observations make it possible to fit the differential absolute
  magnitude distribution $dN/dH$ of the CCs from $5<H_r<13,$ and infer the
differential mass distribution $dN/dM$ across $\ge5$ orders of magnitude in
mass. We find $dN/dM$ well fit by a lognormal distribution, and equally well by
a generalized $\Gamma$ distribution or a double power law. 
The maximum fraction of the total CC mass per log interval in $M$ is in bodies near 75~km diameter, or $\approx10^{-7.5}M_\oplus.$ Extrapolation of the fitted $dN/dH$ functions to $H_r>13$ is unwise, as the different analytic forms diverge. It remains unclear if $dN/dH$ turns over at faint $H.$

The uncertainty in the total mass of
the CC belt is dominated by uncertainty in the relation between $M$ and $H$.   A
calibration using CC binaries suggests a total CC mass of 
1.7--2.7$\times10^{-3}\,M_\oplus.$
A trend toward lower density and/or higher albedo for
smaller bodies may be present in the data, and would lower the estimated total
CC mass.

%These results serve as a target for theories of planetesimal formation.
Qualitative comparison of the derived mass distribution to the results of numerical simulations of the streaming instability (SI) suggest the simulations produce $dN/dM$ distributions that are more sharply peaked, and steeper at the bright end, than the CCs.  Such differences could be ascribed to inhomogeneous formation conditions in the classical belt that are not yet included in modeling.  The variety and uncertainty of $dN/dM$ derived from  state-of-the-art SI simulations currently preclude any definitive test of the SI hypothesis.
\end{abstract}

%% Keywords should appear after the \end{abstract} command. 
%% The AAS Journals now uses Unified Astronomy Thesaurus (UAT) concepts:
%% https://astrothesaurus.org
%% You will be asked to selected these concepts during the submission process
%% but this old "keyword" functionality is maintained in case authors want
%% to include these concepts in their preprints.
%%
%% You can use the \uat command to link your UAT concepts back its source.
\keywords{\uat{Trans-Neptunian objects}{1705} --- \uat{Planetesimals}{1259} --- \uat{Planet Formation}{1476}}

\section{Introduction} 
The cold classical (CC) trans-Neptunian objects (TNOs) are believed to be the only population of bodies in our solar system that have experienced neither severe collisional grinding nor gross dynamical displacement under the influence of the giant planets.  As such they offer a unique \textit{in situ} laboratory to study the stage of planetary evolution when bodies of scales up to $\approx100$~km have formed, but not yet been accreted into larger bodies nor cleared by them.  This population will not be detectable in extrasolar systems for the foreseeable future.  The CC's are our best tool for constraining theories of the formation of 1--100-km-scale planetesimals from proto-planetary nebulae.

The most directly accessible physical property of the CC population is their differential distribution $dN/dH$ of their absolute magnitudes $H,$  which we shall refer to by the usual (but incorrect) title of their ``luminosity function'' (LF).  Under the assumption of spherical bodies with uniform geometric albedo $q,$ $H$ is related to the diameter $D$ as $D \propto 10^{-0.2H}.$  Under the further assumption of uniform bulk density $\rho,$  the mass has $M\propto 10^{-0.6H}.$  The LF is thus a proxy for the size and mass distributions of the bodies, $dN/dD$ and $dN/dM.$

In this work, we use observational data sensitive to CCs with absolute $r$-band
magnitudes as faint as $H_r=13$ to constrain the CC LF.  Since the brightest
known CC (2014~TD$_{86}$) has $H_r\approx 5,$ the observations now span 8
magnitudes of $H$, corresponding to a factor of $\gtrsim10^5$ in mass.  Note
that this span in mass is much larger than the variation in inferred mass due to
plausible variation in bulk density $\rho$, geometric albedo, and shape, so we are justified in taking the broad-range distribution of $H$ as an indicator of the size and mass distributions (see Section~\ref{sec:H2M} for details of the $H$ to $M$ conversion).

Theoretical guidance for the expected functional form of the CC LF is sparse.  A scale-free physical process would typically yield a power-law size distribution.
The first investigation of the CC LF to extend beyond the ground-based limit of $m_r\approx26$~mag was the Hubble Space Telescope survey of \citet{bernstein2004size}. They found that a single power law in $D$ or $M$ fails to fit the observable size distribution, but that ``rolling power laws''---which are lognormal (LN) distributions of size/mass---or a double or broken power law (DPL) can both describe the data (formulae for all functional forms are in Section~\ref{sec:results}).
The DPL reflects the existence of one preferred scale in the formation model, while a LN distribution is the central limit theorem's destination when the TNO size is the product of (powers of) multiple variable initial conditions.

\citet{fraser2014absolute} fit DPL models of $dN/dH$, adding the denser sampling of the ground-based regime provided by the Canada-France Ecliptic Plane Survey \citep{cfeps} and the Subaru data of \citet{fuentes08}.
Much larger controlled sampled of CCs were obtained by the ground-based Outer Solar System Origins Survey \citep[OSSOS,][]{bannister_ossos_2018} and Dark Energy Survey \citep[DES,][]{des2}.  

The emergence of the streaming instability \citep[SI;][]{youdin2005streaming}   and subsequent cloud collapse as a leading theory for the formation of km-scale bodies motivated consideration of a power law cumulative mass distribution with exponential cutoff at high masses [truncated cumulative power-law, or ``TCP'' model, \eqq{eq:ggcdf1}]. While there are no analytic derivations of this particular form, it is seen to describe cloud masses in numerical simulations of SI \citep{johansen2011planetesimal,schafer2017initial,abod2019mass,li_demographics_2019}. An analytic prediction of $dN/dM$ is given by \citet{GerbigLi}.  They assume that the SI or some other mechanism concentrates the pebbles to a high enough surface density to become unstable under the Toomre criterion and resistant to tidal disruption; then make the ansatz that $dN/dM$ will be proportional to the growth rate at the scale corresponding to the Hill radius of $M.$  The shape of $dN/dM$ is largely determined by a dimensionless parameter combination $Q_p,$ with collapse occurring for $Q_p<1.$  Values of $Q_p$ closer to zero yield broader $dN/d\log M$ distributions.  Other individual initial conditions set the mass of peak $dN/dM.$ The analytic predictions are consistent under a K-S test with the masses of $O(100)$ clumps produced in their numerical simulations.

\citet{JJ} combined the OSSOS CC detections with HST and other ground-based discoveries to demonstrate compatibility of the CC LF with the SI-inspired TCP functional form.
More recently, \citet{deepV} find that LN, DPL, and TCP models of the CC LF can all satisfactorily fit preliminary yields of the DECam Ecliptic Exploration Project (DEEP).

\citet{mars} report TNO discoveries using near-infrared (NIR) imaging from the James Webb Space Telescope (JWST) that substantially exceeds both the depth and area of the 2004 HST survey.  A reanalysis of the HST images by \citet{kevin} extends their limiting magnitude and yields 2 new TNO discoveries.  Together, these yield a $\approx7\times$ increase in the number of detected CCs beyond the ground-based limit.

In this work, we essentially repeat the analysis of \citet{JJ} with stronger constraints on the faint-end behavior of the CC LF, and extend it to a broader set of LF models.  We aim to find the simplest model (fewest free parameters) which successfully fits the current observational knowledge of the CC LF, and ask which salient features of the derived LF are independent of the choice of fitting function. This sharpened knowledge of the true CC LF will serve as a target for physical modeling of the formation of planetesimals in the primordial conditions of our outer Solar System. 

The review of planetesimal formation theory by \citet{SimonReview} includes description of the SI and graphical comparison to the CC LF of \citet{JJ}.  While this paper was under review, a quantitative comparison of our derived CC LF (and earlier fits to the CC LF) to numerical SI simulations was released by \citet{Hall2026}.

\section{Sample selection}

To constrain the CC LF across the full range of detected members, we combine likelihoods for the results of four different well-defined samples, summarized in Table~\ref{tab:surveys}.  These surveys, and their detection limits in their native filter systems, are:
\begin{itemize}
    \item The faintest and smallest known TNOs were discovered in the JWST observations of \citet{mars}, which is $>50\%$ complete for magnitude $<28.9$ in the NIRCAM F150W2 filter.
    \item Less deep but useful are the detections from the HST survey of \citet{bernstein2004size}.  A recent reanalysis by \citet{kevin} pushed the 50\% detection limit to 29.2~mag in the ACS F606W STMAG system.
    \item Much more extensive data is available at depths accessible to wide-field ground-based cameras.  We use the catalog from OSSOS survey conducted at the Canada-France-Hawaii Telescope (CFHT) \citep{bannister_ossos_2018}.  The 7 OSSOS fields reach 40\% completeness at $r$-band magnitudes of 24.1--25.4.
    \item Following \citet{JJ}, we assemble a ``Bright'' sample by extracting all objects with CC dynamics from the MPC database, and assuming that this sample is nearly complete to some magnitude limit, largely through the efforts of the Pan-STARRS surveys \citep{ps1}.   We generate two different Bright samples, one limited by absolute magnitude $H_r,$ and another limited by apparent magnitude $m_r$ at opposition.
\end{itemize}

\begin{deluxetable}{lcccccl}
\tablewidth{0pt}
\tablecaption{Observations used \label{tab:surveys}}
\tablehead{
\colhead{Survey} &
\colhead{Selection} &
\colhead{$N_{CC}$} &
\colhead{Area} &
\colhead{Mag limit\tablenotemark{a}} &
\colhead{Assumed color} &
\colhead{Reference}}
\startdata
Bright-app & Eqns~(\ref{eq:cc1}) & 17 & full & $m_r<21.85$ & $r-w=-0.05$ & MPC\tablenotemark{b} \\
Bright-abs & Eqns~(\ref{eq:cc1}) & 13 & full & $H_r<5.50$ & $r-w=-0.05$ & MPC\tablenotemark{b} \\
OSSOS & Eqns~(\ref{eq:cc1}) & 233 & 91~deg$^2$ & $m_r<24.1\ldots25.4$ & \nodata & \citet{bannister_ossos_2018} \\
HST & Eqns~(\ref{eq:cc2}) & 5 & 0.019~deg$^2$ & $m_r<28.95$  & $r-F606W=-0.3$ & \citet{kevin} \\
JWST &Eqns~(\ref{eq:cc2}) & 15 & 0.039~deg$^2$ & $m_r<29.22$  & $r-F150W2= 0.37$ & \citet{mars} \\
\enddata
\tablenotetext{a}{$r$-band magnitude at which selection function falls to 40\%.}
\tablenotetext{b}{Only one of the Bright samples is included for any given analysis.  Selections are made from the distant-object files at \url{https://www.minorplanetcenter.net/data}.}
\end{deluxetable}

In order to evaluate the likelihood functions for the survey outputs described in Section~\ref{sec:likeli}, we need the following for each survey field:
\begin{itemize}
\item The distribution of the survey area as a function of invariable latitude, $dA/db_{\rm inv}.$ Equivalently we may use the full surveyed solid angle $A$ and the range of $b_{\rm inv}$ spanned by the survey field, since all of the fields are nearly rectangular.
\item The detection efficiency $p_{\rm det}(m_r)$ within this area as a function of apparent $r$-band magnitude.  We truncate each field's detection efficiency at the $m_r$ for which $p_{\rm det}$ drops to 40\% of its peak (bright-end) value, and discard detections below this limit.  This reduces the results' sensitivity to the details of the detection fall-off.
\item The solar elongation of the survey field at the time of discovery observations.
\item The list of detected sources meeting the CC criteria, and the values of the barycentric distance $d_{\rm bary},$ plus $H_r, m_r,$ and $b_{\rm inv}$ at the time of detection.
\end{itemize}

%Table~\ref{tab:fields} gives the required information on each field or sample we use.  Table~\ref{tab:tnos} gives the requisite information for all TNOs passing CC sample selections.

We standardize magnitudes from all surveys into the Sloan $r$ band.  All magnitudes are on the AB system except for $R$ band (Vega system) and the HST F606W band (STMAG system).  The following subsections describe the extraction and standardization of CC samples from all 4 catalogs.  We apply the phase correction of \citet{Bowell1989} used by \citet{JJ} to all detections in converting apparent to absolute magnitudes, regardless of wavelength. The correction vs phase angle $\alpha$ is given by

\begin{equation}
\Delta H = \log((1-G)\,\Phi_1 + G\,\Phi_2)
\end{equation}
\noindent where 
\begin{eqnarray}
    \Phi_1 & = & \exp\left(-3.33 \tan\left(\frac{\alpha}{2}\right)^{0.63}\right) \\
    \Phi_2 & = & \exp\left(-1.87 \tan\left(\frac{\alpha}{2}\right)^{1.22}\right)
\end{eqnarray} 
\noindent
and G=0.1.

\subsection{Definition(s) of the cold classicals}
We adopt the same criteria for CC membership as \citet{JJ}:
\begin{equation}
    \begin{array}{c}
 42.7\,\textrm{au}  < a < 47.8\,\textrm{au}, \\
 i_{\rm free}  < 4\degr, \\
 e  < 0.15, \\
\textrm{No identified mean-motion resonance.}
\end{array}
\label{eq:cc1}
 \end{equation}
Free inclinations $i_{\rm free}$ for many TNOs are derived by \citet{yukun} and an extension of that work (Y. Huang, private communication).  This includes nearly all of the TNOs in OSSOS, or listed as $H<6.5$ by the MPC, that meet the $a$ and $e$ criteria above, have non-resonant dynamics, and inclination $i_{\rm inv}<6\degr$ with respect to the invariable plane.  We can thus apply these criteria to our OSSOS and Bright samples. 

For the objects discovered by HST and JWST, however, the $\approx10$-day arcs are insufficient to measure the line-of-sight velocity component of the discoveries' state vectors, leaving a degeneracy between the orbital elements $a$ and $e,$ so we cannot apply Eqns.~(\ref{eq:cc1}) to select CCs. We instead designate as CC those discoveries satisfying
\begin{equation}
\begin{array}{c}
 40\,\textrm{au}  < d_{\rm bary} < 50\,\textrm{au},\\
 i_{\rm inv}  < 4\degr, \\
 e_{\rm min}  < 0.15, 
\end{array}
\label{eq:cc2}
\end{equation}
since $i_{\rm inv}$ and the barycentric distance at discovery $d_{\rm bary}$ are well determined by the short arcs, and a lower bound $e_{\rm min}$ on the eccentricity can also be obtained.  We will ignore as unimportant the difference between heliocentric and barycentric distances.

The impact of this mismatch in selection criteria can be assessed using the TNO catalogs for both OSSOS and for the discoveries from the DES \citep{des1}, since both surveys obtained full orbital elements for samples of $\approx{800}$ TNOs.  In this discussion will refer to TNOs passing the dynamical criteria of (\ref{eq:cc1}) as ``true CCs'' and those passing Eqns.~(\ref{eq:cc2}) as simply ``cold.''

\begin{figure}
\includegraphics[width=0.55\textwidth]{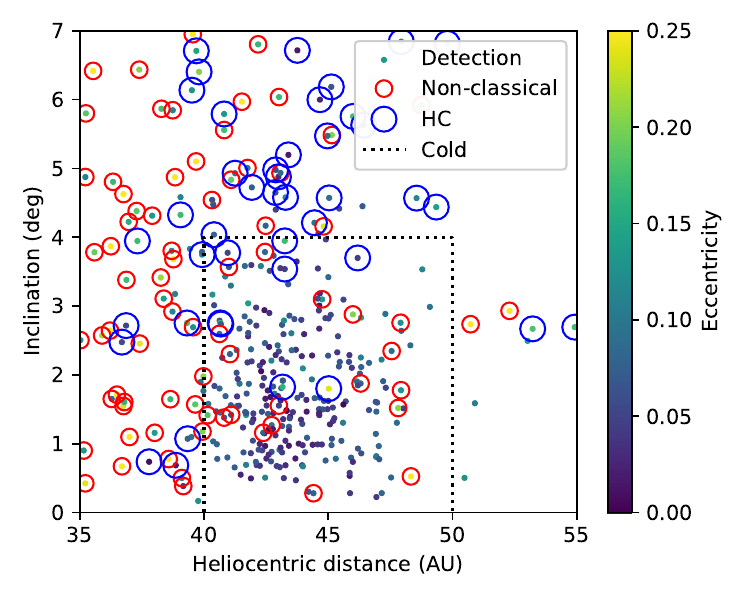}
\includegraphics[width=0.44\textwidth]{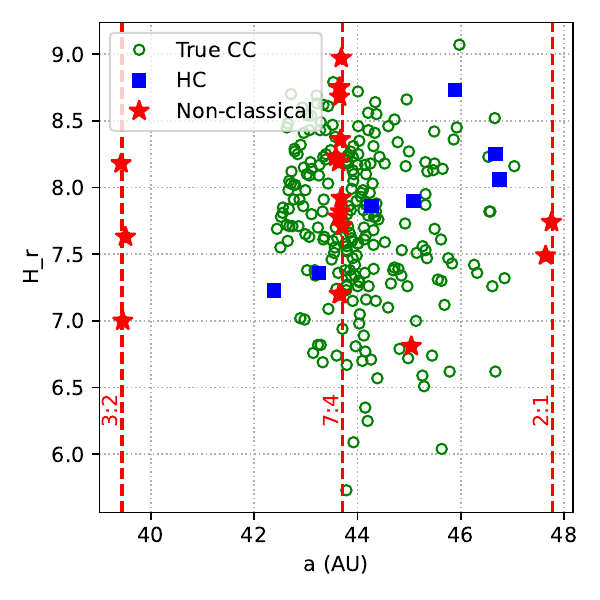}
\caption{\textit{Left:} TNOs detected by OSSOS are plotted at their discovery distance and invariable inclination.  The ``cold'' sample is defined to be those lying within the dotted box.  The points are circled if they are not classical or if they are hot classicals (HC) according to their $i_{\rm free}$ or $e$ values, so that uncircled points are the ``true CCs.''  \textit{Right:} the cold OSSOS TNOs are replotted in the $(a,H_r)$ plane, with mean-motion resonances marked.  Most of the cold sample that are not true CCs are in these resonances. See text for further discussion.}
\label{fig:ossos}
\end{figure}

The left plot of Figures~\ref{fig:ossos} marks the location of  OSSOS-detected TNOs in the space of $(d_{\rm bary}, i_{\rm inv}).$ The dotted boxed area is the selection region for cold objects.  TNOs with resonant or scattering dynamics are circled in red.  Those with classical dynamics but $i_{\rm free}>4\degr$ or $e>0.15,$ \ie\ the hot classicals (HCs), are circled in blue.  The $\approx250$ uncircled points are the true CCs.  Of the true CCs, 7.4\% are missed by the cold criteria.  Most of the missed true CCs are on excursions outside of the 40--50~\au\ range.  Of the cold objects falling in the box, 9.2\% are false true CCs, \ie\ interlopers that are not true CCs.  The right panel plots the OSSOS cold sample in the plane of $(a,H_r),$ where we can see that most of the interlopers are in 3:2, 7:4, or 2:1 mean motion resonances with Neptune.  

We repeat the same test using the TNOs detected by the DES.  The 96 DES TNOs assigned as cold have missed 9.8\% of the true CCs, and contain 13\% interloping resonant TNOs---half of which have $e>0.15$ and might be excluded from the cold classification were they in the JWST or HST fields.

In short, both OSSOS and DES have cold samples that exclude 8--10\% of true CCs, but also replace these with almost identical numbers of non-CC interlopers.  These discrepancies will not affect our fitting of a LF across the full $H$ range, since there are only 20 CCs in the HST$+$JWST samples, so a 10\% mismatch in normalization to the OSSOS and Bright samples would be well below the Poisson noise.  In addition, many of the interlopers are from resonances that lie within the CC belt, which \citet{des2} show are a better color match to the CC population than the HC population, hence may actually be swept-up CCs drawn from the CC LF.  Furthermore, within the precision of the OSSOS dataset, the size distributions of dynamically excited objects is indistinguishable from that of the CCs in the range $6\lesssim H\lesssim9$ \citep{des2,petitLF}. If this holds to fainter $H,$ then the interlopers would not bias our estimated LF shape at all.  \citet{mars} in fact find compatible LF slopes (vs $m_r$) for the cold and hot TNOs across the range probed by the JWST data.

It would be possible to make the cold selection on all of the samples, instead of using true CC dynamical selections for the Bright and OSSOS populations.  It is, however, known since at least \citet{bernstein2004size} that the dynamically excited populations greatly outnumber the cold population at the largest sizes.  We would therefore run significant risk of admitting interlopers into the Bright and OSSOS cold samples that would greatly bias the bright-end behavior of the inferred CC LF.  We judge it better to be able to rigorously exclude non-CCs from the bright end, and absorb the possible $<10\%$ amplitude shift faintward of the OSSOS sample from having different selections in the bright and faint surveys we use.

\subsection{JWST sample}
\label{sec:jwst}
We adopt the detections and efficiency formulations of \citet{mars}.  To obtain a color value between the observed NIRCAM F150W2 filter and the target $r$ filter, we integrate the NIRSpec spectra of TNOs taken by the DiSCo program \citep{disco1,disco_cliff2} across the two filter response functions.  It is necessary to extrapolate the DiSCo spectra into the visible to span the $r$ filter; this is done by adopting the assumption that surface reflectance is a linear function of $\lambda$ across the visible with some slope $S$ \citep[\eg][]{sheppard_color}.  Using the $S\approx15$ slope measured at the blue end of the DiSCo spectra (0.71--0.90 \micron), the average  
result across 5 DiSCo spectra that are classified as ``Cliff 2 cold classical'' types---to which all CCs in the DiSCo sample belong---is $r-\textrm{F150W2}=0.37$~mag.  
If instead we adopt $S\approx30$ as more typical of optical observations of CCs \citep{sheppard_color}, the result is $r-\textrm{F150W2}=0.49.$
We assign $r-\textrm{F150W2}=0.37$~mag to all of the JWST detections.
In Section~\ref{sec:colorcheck} we show that our results are insensitive to changes of 0.2--0.3~mag in this assumed color.

Note that $r-\textrm{F150W2}=0.37$~mag differs appreciably from the $\approx1.0$~mag estimated by \citet{mars} based on spectroscopy of Arrokoth by the New Horizons spacecraft \citep{grundy2020color}, which may be subject to a combination of Arrokoth being atypical, and calibration errors across the different instruments aboard New Horizons (Howett et al, in prep.).

The \citet{mars} catalogs are divided into two mutually exclusive selections.  Sample ``JWSTA'' consists of all those TNOs found in all 3 observing epochs using standard procedures, and holds 11 sources passing the ``cold'' criteria in \eqq{eq:cc2}, once we exclude the targeted \gk\  (which happens to be in the OSSOS sample). Sample ``JWSTB'' contains TNOs detected in exactly two epochs of the search, at $S/N>8$ both times, and holds 4 additional cold discoveries.  

The JWST observations cover 0.05~deg$^2$ at an invariable latitude of -0.46\degr.  We ignore variation of $b_{\rm inv}$ across the small HST and JWST fields.  For any given TNO elements $(a,e,i,\omega)$, we can isolate regions in the $(\Omega, M)$ space for which the TNO would land on a NIRCAM SW detector in all three epochs, which defines a geometric sky area $A_3.$ The value of $A_3$ is found (by sampling orbits) to be, on average, 0.039~deg$^2,$ with weak dependence on the orbital elements. We adopt this as the nominal survey area. The selection functions for JWSTA and JWSTB are constructed as follows, using combinatoric logic and the per-epoch selection functions $p_5$ and $p_8$ derived from source injections:
\begin{align}
    p_A(m) & = p_5^3(m) \\
    p_B(m) & = (A_2/A_3) p_8^2(m) + 3 p_8^2(m) [1-p_5(m)], \\
    p_j(m) & \equiv \frac{p_{\rm bright}}{2}\, {\rm erfc}\left(\frac{m-m_{j0}}{w_j}\right),
\end{align}
where $m$ is the apparent native F150W2 magnitude, and the parameters of the single-epoch functions $p_j$ [$j\in (5,8)$] are: $p_{\rm bright}=0.96, m_{50}=28.92, m_{80}=28.74, w_5=0.61,  w_8=0.40.$  The effective geometric area for TNOs that land on a JWST detector in exactly two of the three epochs is $A_2,$ with  $A_2/A_3=0.42.$  Referred to $r$ band, the limiting apparent magnitudes (40\% completeness) are 29.0 and 29.2 for JWSTA and JWSTB, respectively.  These correspond to limits of $H_r<12.8$ and 13.0 for objects near the 40~AU inner bound of the cold sample.

\subsection{HST sample}
\label{sec:hst}
\citet{bernstein2004size} report a search for TNOs over $A=0.019$~deg$^2$ of sky on the invariable plane from images from the HST ACS instrument in the F606W filter.  One detection was a targeted relatively bright TNO, which we ignore, and 3 other objects were discovered.  \citet{kevin} reanalyzed these images, stacking the observing epochs to obtain slightly fainter detection limit, adding 2 additional detections.  All 5 discoveries satisfy the cold criteria in \eqq{eq:cc2}.  \citet{kevin} gives the detection probability in the native F606W STMAG system as
\begin{equation}
    p_{\rm det}(m) = \left[ 1 + \exp\left(\frac{m-29.21}{0.11}\right)\right]^{-1}.
\label{eq:phst}
\end{equation}
The synthetic photometry process with extrapolated DISCO Cliff 2 spectra described in Section~\ref{sec:jwst} is applied to yield a mean $r-{\rm F606WST}=-0.3,$ which we apply to all detections.  This leads to a detection limit of $H_r<12.7$~mag for sources at $d_{\rm bary}=40$~\au.

\subsection{OSSOS sample}
\citet{bannister_ossos_2018} provide dynamical clasifications, orbital elements, $r$-band magnitudes, and phase-corrected $H_r$ estimates for over 800 TNOs detected by the OSSOS survey.  Adding $i_{\rm free}$ estimates from \citet{yukun}, we are able to apply the ``true CC'' criteria of \eqq{eq:cc1} to yield 233 CCs spread among 7 different rectangular sky fields, or ``blocks.''  The limiting $r$-band magnitudes of these fields range from $24.1<r_{\rm max}<25.2$~mag; completeness functions $p_{\rm det}(m_r)$ are given by \citet{bannister_ossos_2018} for each block.  

Each block spans between 2\degr\ and 4\degr\ of invariable latitude, so we will integrate over these ranges when evaluating likelihoods.  All OSSOS imaging has $|b_{\rm inv}|<5\degr.$

We note that another 100 CCs were discovered in the DES survey reported by \citet{des1}.  We do not include them as constraints on the CC LF since we already have $>2\times$ more CCs from OSSOS over similar magnitude range, and the selection process for DES  (involving linkage over 6 years, during which CCs can drift in/out of the field) is more complex.

\subsection{The brightest CCs}
\label{sec:bright}
Essentially the entire sky region within $\pm4\degr$ of the invariable plane has been searched for TNOs, most notably by the Pan-STARRS surveys \citep{ps1} that cover nearly this entire region.  Following \citet{JJ} we try to estimate an absolute and apparent limiting magnitude brighter than which no new CC's have been discovered since the bulk of PS discoveries were reported to the MPC.  We take the year of the Minor Planet Center (MPC) principal designation of the source as its year of discovery.

From the full listing of distant objects from the MPC (retrieved 25 May 2026), we select those with $a,i,e,$ values that could plausibly lead to classification as true CCs under \eqq{eq:cc1}, and $H_{\rm MPC}<7.$ Then we will examine this MPC subset to find the $m_r$ and $H_r$ values for which no additional CC discoveries have been made since 2015, when most Pan-STARRS TNOs were discovered.
First, we attempt to obtain more accurate $H_r$ estimates than can be derived from the MPC's own $H$ estimates (nominally in $V$ band, but composed of photometry with diverse filters and quality).  To do so we download all observations reported to the MPC for each candidate CC.  An $H_r$ value is estimated through the following sequence:
\begin{itemize}
\item If observations in the PS $w$ filter are recorded (filter codes ``w'' or ``Pw'' in the MPC ADES format), we convert each to an $H_w$ value given the phase and distance corrections appropriate to the observation, and convert the median value to $H_r$ assuming $H_r-H_w=-0.05$~mag. This is the color obtained through synthetic photometry of the DiSCo spectra with linear extrapolation across visible wavelengths as described in Section~\ref{sec:jwst}. The value changes by only 0.02~mag if the linear slope $S$ vs wavelength is changed from 15 to 30.  This weak dependence is because the $w$ passband is close to centered on the $r$ band.
\item If no $w$ observations are available, we search for observations reported in the $r$ filter (this includes Rubin observatory measurements reported as ``Lr'' filter).  If present, they are converted to $H_r$ values and the median value is assigned to the TNO.
\item In the absence of $w$ or $r$ observations, we seek any $R$-band data (presumably in Vega rather than AB magnitudes), take the median implied $H_R,$ and assume $H_r-H_R=0.25$~mag.
\item Only one potentially included TNO, 2011~LJ$_{29}$, has no $w,r,$ or $R$ photometry listed.  We located it on archival $r$-band images from the Dark Energy Camera on the Blanco telescope, and measured $H_r=5.74\pm0.10,$ which ends up excluding this from the selection.
\end{itemize}
The $H_r$ and $m_r$ values assigned to the Bright TNOs should be considered uncertain by at least 0.1~mag, due to sparse light-curve sampling, uncertain color corrections, and low-quality measurements submitted to the MPC. Since there is little structure in the LF on 0.1-mag scales, these uncertainties are not significant to our results.

We also exclude from consideration any TNO assigned an orbit quality code $U$ of \texttt{E}, meaning that the observed arc is too short to determine an orbit.  This excludes one object, 2004~EO$_{95},$ that would have barely passed the criteria for the Bright samples if its elements were taken at face value.  It has only a 1-day arc from 2004 and has never been recovered, and could very well be spurious.

For each of the candidate Bright CCs from the MPC catalog, we now have orbital elements, an estimated $H_r,$ and a year of discovery.  We assign an apparent magnitude $m_r,$ distance $d_{\rm bary},$ and invariable latitude $b_{\rm inv}$ that the TNO would have had if observed at opposition in the year of its MPC designation (\ie\ zero phase correction).

Nearly all of the MPC Bright CC candidates have $i_{\rm free}$ reported by \citet{yukun}, so we can apply the true CC cuts in \eqq{eq:cc1} to them.  Of the objects without tabulated $i_{\rm free},$ none end up passing all the other criteria.

\begin{figure}
    \plottwo{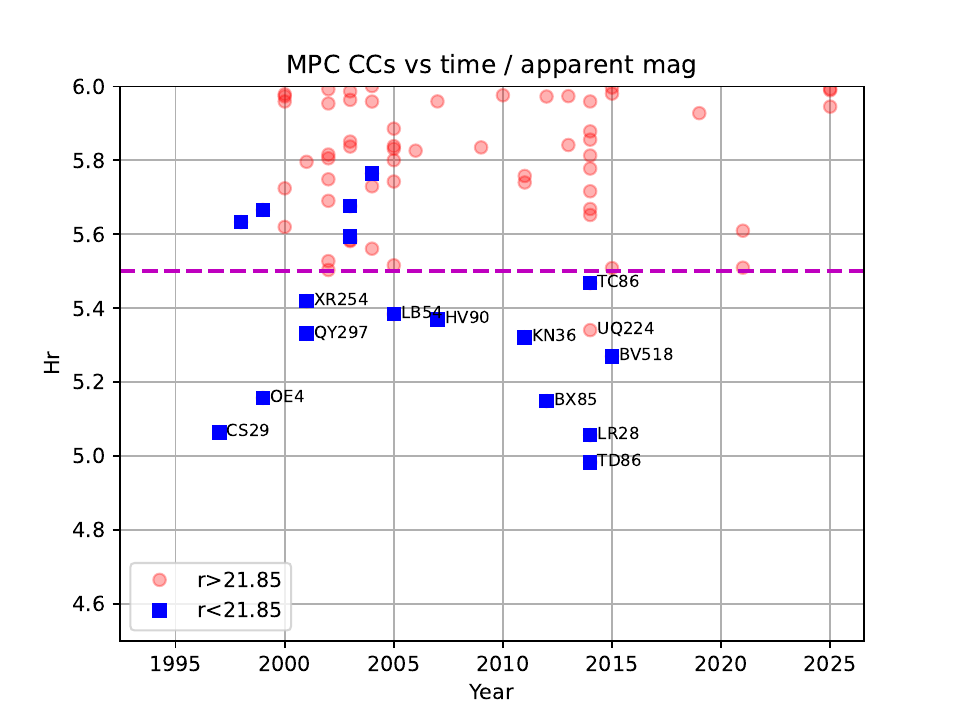}{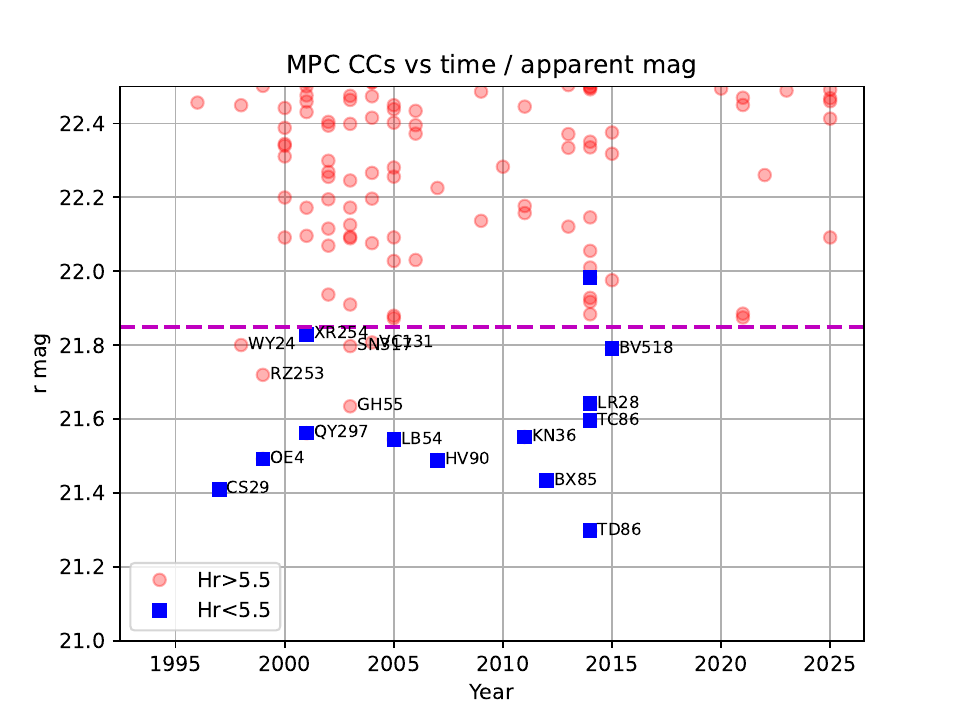}
    \caption{Each panel plots the year of discovery of bright CC TNOs known to the MPC on the $x$ axis, with absolute (left) and apparent (right) $r$ magnitude on the $y$ axis.  Since the bulk of Pan-STARRS discoveries were reported in 2014 and 2015, no new discoveries at $H_r<5.50$ or $m_r<21.85$ (magenta lines) have been reported.  We approximate the MPC sample of CCs as complete in those regimes.}
    \label{fig:mpc}
\end{figure}

We can now proceed to make a judgment of the absolute and apparent magnitudes at which the MPC sample might be considered nearly complete over the CCs.  Figure~\ref{fig:mpc} illustrates that no new CCs have been discovered since 2015 that have $H_r>5.50$ or opposition $m_r>21.85.$ As it is not obvious which if either of the absolute or apparent selections is truly complete, we define a ``Bright-abs(olute)'' and ``Bright-app(arent)'' sample as those meeting each cut.  There are 13 TNOs in Bright-abs, with (523745) 2014~TD$_{86}$ at $H_r=4.98$ the brightest.  There are 17 in Bright-app (12 in common with Bright-abs), the brightest again 2014~TD$_{86}$ at $m_r=21.30.$ The union of the two Bright samples is listed in Table~\ref{tab:bright}.

\begin{deluxetable}{ccc|ccc}
\tablecaption{CC TNOs used in the Bright samples \label{tab:bright}}
\tablewidth{0pt}
\tablehead{
\colhead{Designation} &
\colhead{$m_r$\tablenotemark{a}} &
\colhead{$H_r$\tablenotemark{b}} &
\colhead{Designation} &
\colhead{$m_r$\tablenotemark{a}} &
\colhead{$H_r$\tablenotemark{b}} }
\startdata
2014 TD$_{86}$ & 21.30 & 4.98 & 2007 HV$_{90}$ & 21.49 & 5.37 \\
2014 LR$_{28}$ & 21.64 & 5.06 & 2005 LB$_{54}$ & 21.54 & 5.38 \\
1997 CS$_{29}$ & 21.41 & 5.06 & 2001 XR$_{254}$ & 21.83 & 5.42 \\
2012 BX$_{85}$ & 21.43 & 5.15 & 2014 TC$_{86}$ & 21.60 & 5.47 \\
1999 OE$_{4}$ & 21.49 & 5.16 & 2003 GH$_{55}$ & 21.63 & 5.59 \\
2015 BV$_{518}$ & 21.79 & 5.27 & 1998 WY$_{24}$ & 21.80 & 5.63 \\
2011 KN$_{36}$ & 21.55 & 5.32 & 1999 RZ$_{253}$ & 21.72 & 5.67 \\
2001 QY$_{297}$ & 21.56 & 5.33 & 2003 SN$_{317}$ & 21.80 & 5.68 \\
2014 UQ$_{224}$ & 21.98 & 5.34 & 2004 VC$_{131}$ & 21.81 & 5.76 \\
\enddata
\tablenotetext{a}{Estimated $r$-band apparent magnitude at opposition in year of designation.  The Bright-app sample contains those with $m_r<21.85.$}
\tablenotetext{b}{Estimated $r$-band absolute magnitude.  The Bright-abs sample contains those with $H_r<5.50$}
\end{deluxetable}

We will take the Bright-app sample as our baseline for a complete sample over the entire CC population, \ie\ $p_{\rm det}=1.$  We will also demonstrate that instead adopting the Bright-abs sample does not alter our conclusions. %We will also test whether changing to the Bright-abs sample changes any of our conclusions.  It does not.

\section{Distribution function methodology}
\label{sec:likeli}
Our goal is to assign the probability $p(q_{\rm H} | \data)$ of some parameters $q_H$ of a model for the differential distribution $n(H;q_H)\equiv dN/dH_r$ of the total number of CCs, conditioned on the data $\data$ consisting of the collected contents of the Bright, OSSOS, HST, and JWST observational samples.  

Because our surveys select by apparent magnitude $m_r$ instead of $H_r$ (with the exception of Bright-abs), and because our coverage is not complete over the entire $|b_{\rm inv}|<4\degr$ range of invariable latitude that CCs could occupy, we cannot predict the data from an $n(H)$ model alone.  We need a multidimensional model for $dN/dH_r\,db_{\rm inv}\,d\mu,$ where $\mu$ is some indicator of the instantaneous barycentric distance of the CC TNOs. We will adopt the definition
\begin{equation}
\mu = \frac{ \log (d_{\rm bary}/40\,{\rm AU})}{\log (50/40)},
\label{eq:mu}
\end{equation}
which varies from $0<\mu<1$ as we traverse the ``cold'' selection region for the HST and JWST samples.\footnote{About 2\% of OSSOS true CCs have $d_{\rm bary}$ outside the 40--50 \au\ range and hence have $\mu<0$ or $\mu>1.$ Our model distributions $g(\mu)$ extend to this range but are normalized to unit integral in $0<\mu<1.$}  With knowledge of the solar elongation $\beta$ of an observation, specification of $\mu$ determines the geocentric and barycentric distances $d_{\rm geo}$ and $d_{\rm bary}$ as well as the phase $\phi$ and the assumed phase function $G(\phi).$  This allows us to tabulate a function $\Delta_i(\mu)$ for each observed field $i$ that combines the distances and phase factor into the difference between apparent and absolute magnitudes:
\begin{equation}
    m_r = H_r + \Delta_i(\mu).
    \label{eq:mH}
\end{equation}

We will assume that the true instantaneous CC distribution is separable as
\begin{align}
    \frac{dN}{dH_r\, db_{\rm inv}\, d\mu}
    & = N_{\rm tot} n(H_r) f(b_{\rm inv}) g(\mu) 
    \label{eq:nfg}\\
    \int_4^{14} n(H_r)dH_r & = 1 \label{eq:nnorm}\\
    \int_{-4\degr}^{4\degr} f(b) db & = 1 \label{eq:fnorm}\\
    \int_0^1 g(\mu) d\mu & = 1.
    \label{eq:gnorm}
\end{align}
In this case, $N_{\rm tot}$ specifies the total number of CCs between 40 and 50~AU, with $4<H_r<14.$

The expected number of detections in a survey segment spanning solid angle $A$ and invariable latitude $b_-<b_{\rm inv}<b_+$ becomes
\begin{equation}
    \bar N = \frac{A N_{\rm tot}}{360\degr} F_b(q_b) \int dH_r\,  n(H_r;q_H) \int d\mu \,g(\mu;q_\mu) p_{det}\left[H_r + \Delta(\mu)\right],
\label{eq:barN}
\end{equation}
where we have introduced parameters $q_b$ and $q_\mu$ describing the unit-normalized latitude and distance distributions of the instantaneous CC population.  The latitude coverage factor is
\begin{equation}
   F_b(q_b) \equiv \int_{b_-}^{b_+} db\, f(b;q_b).
\label{eq:meanb}
\end{equation}
For the Bright samples, the assumed full-sky coverage means $AF_b = 360\degr.$  For the Bright-abs sample, $p_{\rm det}$ is a step function in $H_r$ so the $\mu$ integral is not needed.

Assuming that the detections are drawn independently from the distribution specified by the model in \eqq{eq:nfg}, then the log-likelihood of the data $\data_i$ in survey field $i$ with detections indexed by $j$ takes the Poisson form
\begin{equation}
    \log \likeli(\data_i|N_{\rm tot},q_m, q_b, q_\mu) =
    -\bar N_i(N_{\rm tot},q_m, q_b, q_\mu)
    + \sum_j \log \left\{ N_{\rm tot} g(b_{{\rm inv},j,};q_b) n(H_{r,j};q_H) f(\mu_{j};q_\mu) p_{{\rm det},i}\left[H_{r,j}+\Delta_i(\mu_j)\right]\right\}
\end{equation}

Formally this expression is in error, because the model $n(H_r)$ and the detection probability $p_{\rm det}(m_r)$ are defined on the \textit{true} magnitudes of the sources, but we evaluate them at the \textit{observed} magnitudes of the detected sources.  The biases we incur from this error are, however, expected to be small, since the $n(H_r)$ function and, usually, the $p_{\rm det}$ function, vary only slightly across the span of the typical magnitude measurement errors ($\approx0.1$~mag).  And, again, with only 20 CCs in the 2 fainter samples, the Poisson uncertainties will dominate any resultant biases.

Summing over all the survey fields (including the Bright sample) yields the total log-likelihood of the data given the model parameters, $\log \likeli(\data | N_{\rm tot},q_H,q_b,q_\mu).$  We define the log-likelihood of a given choice of $q_H$ conditioned on the observations to be
\begin{equation}
L(q_H | \data) \equiv 
\log \,\underset{N_{\rm tot},q_b,q_\mu}{\operatorname{max}} \likeli(\data|N_{\rm tot}, q_H,q_b,q_\mu).
\end{equation}
This is the Bayesian posterior log probability for $q_H$ (up to a constant) assuming uniform priors for the arguments $q_H,$ except that we are taking the \emph{profile likelihood} over the nuisance parameters $(N_{\rm tot},q_b,q_\mu)$ instead of marginalizing over them.  This removes sensitivity of the maximum $L$ location to the assignment of priors to the nuisance variables.  The latitude and distance nuisances are quite strongly constrained by the data, in any case, as we show below.  Furthermore, the solution for the maximizing $N_{\rm tot}$ value is analytic once the other terms are calculated.

\subsection{Geometric models}
For the distribution of CC's in invariable latitude, we adopt a normal distribution with zero mean and a standard deviation $\sigma_b$ as the only element of $q_b.$  The normal is truncated at $b_{\rm inv}=\pm4\degr,$ since both the true CC criteria and the cold criteria exclude higher latitudes.  While the distribution of \textit{inclination} $i_{\rm inv}$ has been found to be consistent with normal for CCs \citep{brown_inc,cfeps,des2}, the observed distribution of \textit{latitude} is less quantified.  The assumed $f(b_{\rm inv})$ function will couple to the inferred LF slope, since the deep surveys are near $b_{\rm inv}=0$ but the shallower surveys span the CC range.  The distribution of $b_{\rm inv}$ \textit{within} the OSSOS sample will, however, serve to constrain $\sigma_b.$  The maximum-likelihood values are within $\sigma_b=(1.35\pm0.02)^\circ$ for all cases examined.

\begin{figure}
    \centering
    \includegraphics[width=0.5\textwidth]{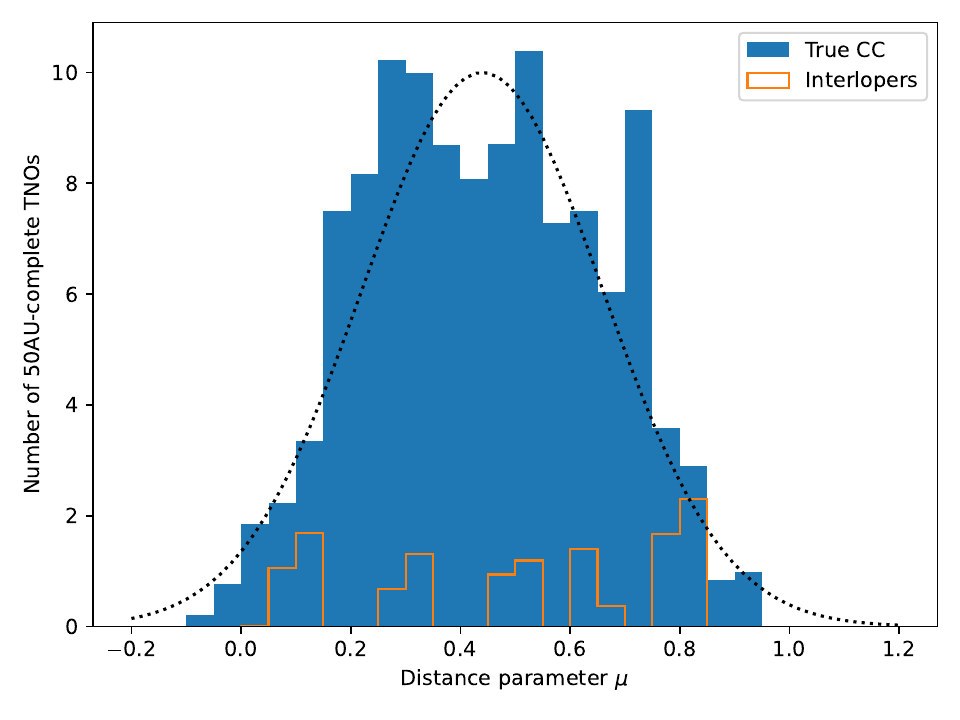}
\caption{The completeness-corrected distribution of the distance parameter $\mu$ is plotted for OSSOS true CCs, and for non-CC interlopers into the cold sample.  We overplot the best-fit normal distribution (dotted line).  Note that $\mu$ is $\log d_{\rm bary}$ scaled so that $0<\mu<1$ corresponds to $40<d_{\rm bary}<50$~\au.}
\label{fig:mu}
\end{figure}

We also adopt for the distance distribution $g(\mu)$ a normal with the mean $\mu_0$ and dispersion $\sigma_\mu$ as the free parameters $q_\mu.$  Figure~\ref{fig:mu} shows a validation of this model using the OSSOS sample.  We debias the observed $\mu$ distribution by taking all objects that would be detectable if placed at $d_{\rm bary}=50$~\au\ ($\mu=1$), and constructing a  histogram of their observed $\mu$ values weighted by $p_{\rm det}\left[H_r + \Delta(\mu=1)\right] / 
p_{\rm det}\left[H_r + \Delta(\mu)\right].$

The dotted curve in Figure~\ref{fig:mu} plots $\mu_0=0.43, \sigma_\mu=0.22,$ which are within 0.01 of the maximum-likelihood values in all fits. This maximum-likelihood Gaussian distribution is seen to give a good visual fit to the debiased OSSOS CC $\mu$ distribution, apart perhaps from some kurtosis present in the data.

\section{Results for flux/size/mass distributions}
\label{sec:results}
In exploring potential forms for the CC LF,
all magnitudes will be assumed to be in $r$ band unless noted otherwise.
When connecting the distribution of apparent magnitude $H$ to the distributions of diameter (size) $D$ and mass $M,$ we denote as \eg\ $M_x$ the mass corresponding to some $r$-band absolute magnitude $H_x$ with the same subscript.

Each functional form for $n(H_r)$ is accompanied by a normalization factor to enforce \eqq{eq:nnorm}.  For convenience we will not include these factors in our definitions.

\subsection{Rolling power law / lognormal distribution}
A single power law distribution for CC TNO sizes has been known to be excluded since at least \citet{bernstein2004size}.
We start here by fitting the ``rolling power law'' model used in that work to the 4 samples, defined by
\begin{equation}
n_{\rm LN}(H_r) \propto 10^{\alpha_1(H_r-H_0) + \alpha_2(H_r-H_0)^2},
\label{eq:roll}
\end{equation}
where $H_0$ is an arbitrarily chosen reference point that we set to $H_r=8$~mag.  This is equivalent to a normal distribution of $H_r,$ or a log-normal distribution of the linear variables $D$ and $M.$  We will refer to this as the ``lognormal'' (LN) model.  There are only 2 free parameters in this model, $q_H=(\alpha_1, \alpha_2)$ (plus the overall normalization).

Maximizing the likelihood yields:
\begin{equation}
    \max \log p_{\rm LN} =
    \begin{cases}
        1572.7 \quad @ \quad(\alpha_1,\alpha_2)=(0.672,-0.094) & \text{(Bright-app)} \\
        1561.5 \quad @ \quad (\alpha_1,\alpha_2)=(0.685, -0.098) & \text{(Bright-abs)} \\        
    \end{cases}
\end{equation}
 We run a Metropolis-Hasting Markov chain (MHMC) over the $(\alpha_1,\alpha_2)$ space for the Bright-app sample to determine uncertainties in the derived values, obtaining a distribution resembling a multivariate Gaussian with $\alpha_1=0.671\pm0.021, \alpha_2=-0.095\pm0.008,$ with correlation coefficient of $r=-0.15$ between the two parameters.  Since the statistical uncertainties are larger than the difference between maximum likelihood (ML) values using absolute vs apparent magnitudes, we take the results as robust to this choice and will restrict further analysis to the Bright-app sample, for clarity.

\begin{figure}
    \plotone{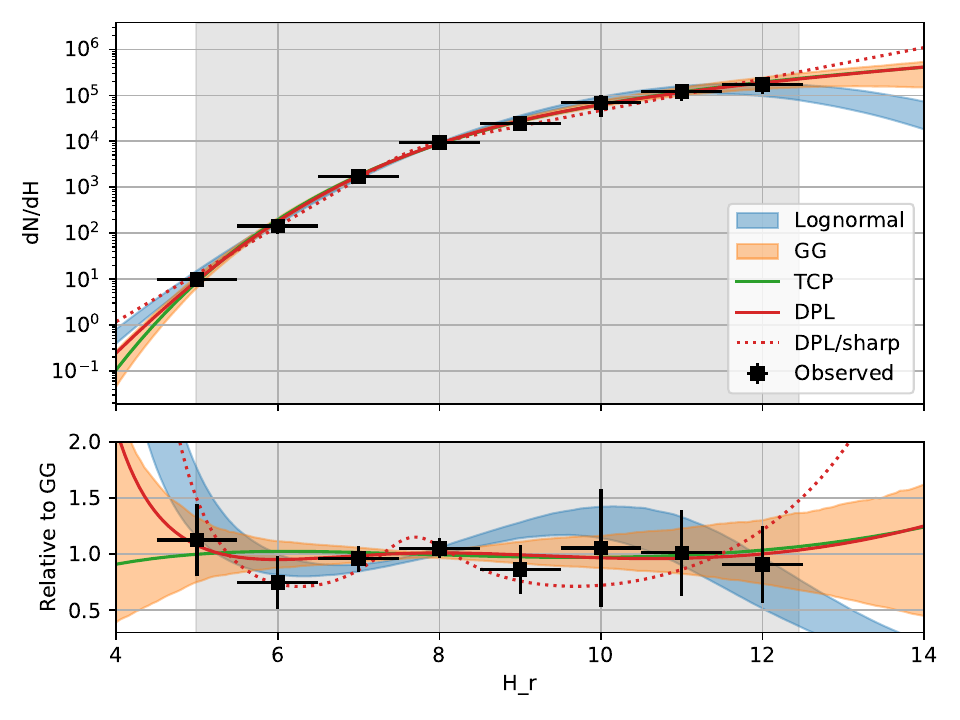}
    \caption{The estimated differential counts of CC TNOs vs absolute $r$ magnitude are plotted.  The curves are from the maximum-likelihood fitted models of the listed functional forms. The shaded regions for the Lognormal and generalized-$\Gamma$ (GG) models bound the 68\% credible region at each value of $H_r.$ The black squares are the best-fit $dN/dH_r$ values, and their independent Poisson uncertainties, obtained from the observations in individual 1-mag bins.  The lower panel shows the results relative to the maximum-likelihood GG model.  [Note that the fits are done to the full detection catalogs, not this binned representation.  And each binned point assumes that the shape of the LF \textit{within} the bin follows the best-fit GG distribution.]}
    \label{fig:lf}
\end{figure}

 The goodness-of-fit of the LN model to the Bright-app data is determined by drawing 200 sets of 270 CC TNO detections (to match the observed number), and distributing them among the survey fields and values of $H_r, \mu,$ and $b_{\rm inv}$ according to the maximum-likelihood LN model and the known detection probabilities.  For each mock survey we find the maximum log-likelihood by the same procedure as  for the real data.  The real data's $\log p_{\rm LN}$ is at the 51st percentile of the simulated samples, indicating that the observations are fully consistent with being drawn from the LN model.  The 68\% credible range for the lognormal model is plotted in blue in Figure~\ref{fig:lf} along with the $dN/dH$ inferred from the observed object counts and survey efficiencies in 1-mag bins of $H_r.$ 

The fitted parameters imply that the $H_r$ distribution reaches a peak at $H_r=11.57\pm0.33$~mag (68\% credible region), indicating near certainty that the LF reaches a peak within the range observed by JWST \emph{if the diameter/mass distributions are truly lognormal.}

\subsection{Generalized Gamma distributions}
A functional form successfully fit to the mass distribution of clumps in simulations of the streaming instability \citep{schafer2017initial, li_demographics_2019} is a \textit{cumulative} distribution that is a power law with an exponential taper at the bright end---the ``tapered cumulative power law'' or ``TCP'' distribution:
\begin{align}
    n(<H_r) & \propto 10^{\alpha (H_r-H_0)} \exp\left[-10^{\beta(H_b-H_r)}\right]
    \label{eq:ggcdf1} \\
    \Rightarrow\qquad n_{\rm TCP}(H_r)=\frac{dN}{dH_r} & \propto 
    10^{\alpha (H_r-H_0)} \exp\left[10^{-\beta(H_b-H_r)}\right]\left[\alpha + \beta\, 10^{-\beta(H_b-H_r)}\right],
    \label{eq:ggcdf}
\end{align}

We choose to avoid the TCP model (except to compare with previous results) because it compresses a wide range of $dN/dH$ functions that flatten or fall at the faint end into a very narrow range of $\alpha$ parameters, and because the conversion to a PDF is necessary to properly calculate a likelihood anyway.
A variant with an easier-to-understand differential distribution is to let the \textit{differential} counts $dN/dH$ take  
the functional form of \eqq{eq:ggcdf1}:
\begin{align}
n_{\rm GG}(H_r)=\frac{dN}{dH_r} & \propto 10^{\alpha (H_r-H_0)} \exp\left[-10^{\beta(H_b-H_r)}\right]
    \label{eq:ggpdf}\\
    \Rightarrow\qquad \frac{dN}{dD} & \propto 
    D^{-5\alpha-1} e^{-(D/D_b)^{5\beta}}, 
    \label{eq:ggpdfD}\\ 
     \frac{dN}{dM} & \propto 
    M^{-5\alpha/3-1} e^{-(M/M_b)^{5\beta/3}}, 
    \label{eq:ggpdfM}
\end{align}
where $\alpha$ is the asymptotic faint-end slope of $\log_{10} dN/dH,$ $H_b$ marks the onset luminosity of the taper, $\beta>0$ is a parameter that governs the rapidity of the suppression, and $H_0$ is again an arbitrary scaling, leaving 3 shape parameters: $q_H = \{\alpha, \beta, H_b\}.$

The distribution of the physical quantities $D$ or $M$ in  \eqq{eq:ggpdfD} or \eqq{eq:ggpdfM} is known as 
the generalized $\Gamma$ distribution. We therefore refer to the model in \eqq{eq:ggpdf} as the ``GG'' luminosity function.  

The two forms are equivalently good fits to the observations;
maximizing $p(q_H | D)$ for each, we obtain:
\begin{align}
\max \log p_{GG} & = 
\begin{cases}
    1574.8 \quad @ \quad (\alpha,\beta,H_b)=(0.012, 0.145, 12.14) & \text{(Bright-app)} \\
    1563.9 \quad @ \quad (\alpha,\beta,H_b)=(0.047, 0.159, 11.37) & \text{(Bright-abs)}
\end{cases}  
\label{eq:pGG} \\
\max \log p_{TCP} & = 
\begin{cases}
    1574.8 \quad @ \quad (\alpha,\beta,H_b)=(0.125, 0.149, 11.95) & \text{(Bright-app)} \\
    1561.9 \quad @ \quad (\alpha,\beta,H_b)=(0.293, 0.241, 8.76) & \text{(Bright-abs)}
\end{cases}  
\label{eq:pTCP}
\end{align}

The data do not distinguish between GG and TCP forms for any model-discrimination statistic.  As seen in lower panel of Figure~\ref{fig:lf}, the green best-fit TCP curve is within a few percent of the best-fit GG value  across the $H_r$ range of interest.  We will prefer the simpler GG form for further analysis.  

The best GG model exceeds the LN model's probability by just $\Delta \log p=2.1$ using the Bright-app sample, similarly for the Bright-abs sample, and has one additional free parameter, offering no significant preference between LN and GG models.  

To test the goodness-of-fit of the GG model, we generate mock samples from the best-fit GG model, and find the maximum $\log p$ value for fitting each sample with a GG model.  We find the data lie at the 47th percentile of the generated $\log p$ values, indicating that the data are fully consistent with arising from the GG model.

\begin{figure}
    \plotone{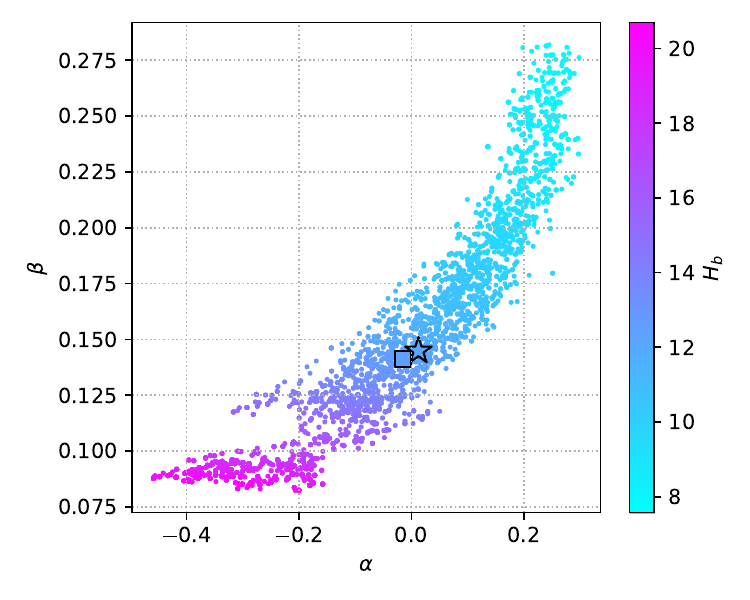}
    \caption{Scatter plot of the MHMC samples from the probability of the generalized $\Gamma$ (GG) distribution for CC luminosity, defined in \eqq{eq:ggpdf}, conditioned on the Bright-app sample.  There is a strong degeneracy between the three free parameters; they scatter around a narrow 1d locus in the 3d space. The star and square mark the maximum-likelihood parameters with the nominal color $r-F150W2=0.37$ and a deviation to 0.6, respectively. Note that the bounds on the faint-end slope $\alpha$ are very model-specific, and this plot demonstrates a degeneracy, not any model-independent determination of $\alpha.$}
\label{fig:ggmh}
\end{figure}

We sample the GG parameters with an MHMC, yielding the narrow range of possible GG LFs depicted as the orange ranges in Figure~\ref{fig:lf}.  We also show in Figure~\ref{fig:ggmh} that there is a strong degeneracy among the 3 parameters of the GG function when fit to the data.
The best-fit GG model has $\alpha=0.012>0,$ meaning that the data do \emph{not} require that the CC LF reaches a maximum.  Further discussion of the results for the faint-end behavior of the derived LF is in Section~\ref{sec:faint}.

\subsection{Double power law (DPL)}
As one additional case, we test whether the data are consistent
with a LF that is asymptotically a power law at both bright and faint ends, defined via:
\begin{equation}
    n_{\rm DPL}(H_r) = \left[ 10^{-\alpha_1\beta(H_r-H_b)}
    + 10^{-\alpha_2\beta(H_r-H_b)}\right]^{-1/\beta}.
    \label{eq:dpl}
\end{equation}
The $n(H_r)$ function is observed to become shallower at fainter magnitudes, so we constrain $\alpha_2>\alpha_1$ and $\beta>0.$  The bright- and faint-end slopes of $\log_{10} n(H_r)$ asymptote to $\alpha_1$ and $\alpha_2,$ crossing at $H_r=H_b$.  Larger values of $\beta$ give sharper transitions between the two asymptotes.  There are 4 free parameters (plus the normalization $N_{\rm tot}$).

The implied function $dN/dD$ asymptotes to power laws with slopes $(-5\alpha_1-1, -5\alpha_2-1)$ at high and low sizes, respectively. Similarly, this is a double power law for $dN/dM$, with slopes $(-5\alpha_1/3-1, -5\alpha_2/3-1).$

The maximum $\log p(q_H | \data)$ obtained for the data under the DPL model are:
\begin{equation}
\max \log p_{\rm DPL} = 
\begin{cases}
    1574.8 & @ \quad (\alpha_1,\alpha_2,H_b,\beta)=(2.21,0.14, 5.87, 0.13) \quad \text{(Bright-app)}\\
    1564.1 & @ \quad (\alpha_1,\alpha_2,H_b,\beta)=(1.45, 0.24, 7.15, 0.38) \quad \text{(Bright-abs)}\\
\end{cases}
\label{eq:pDPL}
\end{equation}    
almost identical to the results from the GG model that has one fewer degree of freedom.  This occurs because the DPL model is almost identical to the GG model for the preferred low values of $\beta,$ as seen in Figure~\ref{fig:lf}. At such a small $\beta$, the logarithmic slope is changing across the full observed magnitude range. 

Enforcing a sudden transition in slope by fixing $\beta=4$ only slightly worsens the fit:
\begin{equation}
\max \log p_{\rm DPL/sharp} = 1573.0 \quad @ \quad (\alpha_1,\alpha_2,H_b)=(1.04, 0.34, 7.66) \quad \text{(Bright-app)},
\label{eq:dpl4}
\end{equation}
only $\Delta \log p=-1.8$ (at one fewer free parameter) worse than the full DPL model.
Hence a sharply broken power law cannot be excluded with current data.  This best fit is plotted as ``DPL/sharp'' in Figure~\ref{fig:lf}. The observations' $\log p$ value lies at the 51st percentile of the $\log p$ values of mock samples drawn from this model, confirming that DPL/sharp is a valid model for the observations.
The parameters in \eqq{eq:dpl4} are within the $2\sigma$ regions of the DPL fits in \citet{bernstein2004size} and \citet{fraser2014absolute}.

\subsection{Robustness to NIR color}
\label{sec:colorcheck}
We substitute an assumed CC visible-NIR color  $r-F150W2=0.6$ for the value $r-F150W2=0.37$ derived herein from DISCO spectra and repeat the maximum-likelihood fits to the models.  For the LN model, the $\log p$ values are decreased by $\approx0.4$ for both Bright-app and Bright-abs cases.  The shifts in the ML parameters are $\Delta\alpha_1\approx-0.002, \Delta\alpha_2\approx-0.003,$ well below statistical uncertainties.

For the GG model, the shift in ML $\log p$ are $<0.01,$ and $|\Delta(\alpha,\beta,H_b)| <(0.07, 0.005, 1),$ with most of the shift along the degeneracy locus as seen in Figure~\ref{fig:ggmh}. We conclude that the uncertainty in the visible-NIR color of CC's is unimportant for our analysis. 

%\citet{li_demographics_2019} fit the mass distribution of simulated SI-induced clumps with a variant of the DPL that includes a sharp cutoff (truncation) above a certain mass.  Truncating our DPL at $H_r$ just below the largest known CC TNO does not significantly alter the $\log p$ values or best-fit parameters, because the best-fit untruncated model predicts an average of only $\approx 2$ additional TNOs brighter than known.  

\begin{figure}
    \includegraphics[width=\textwidth]{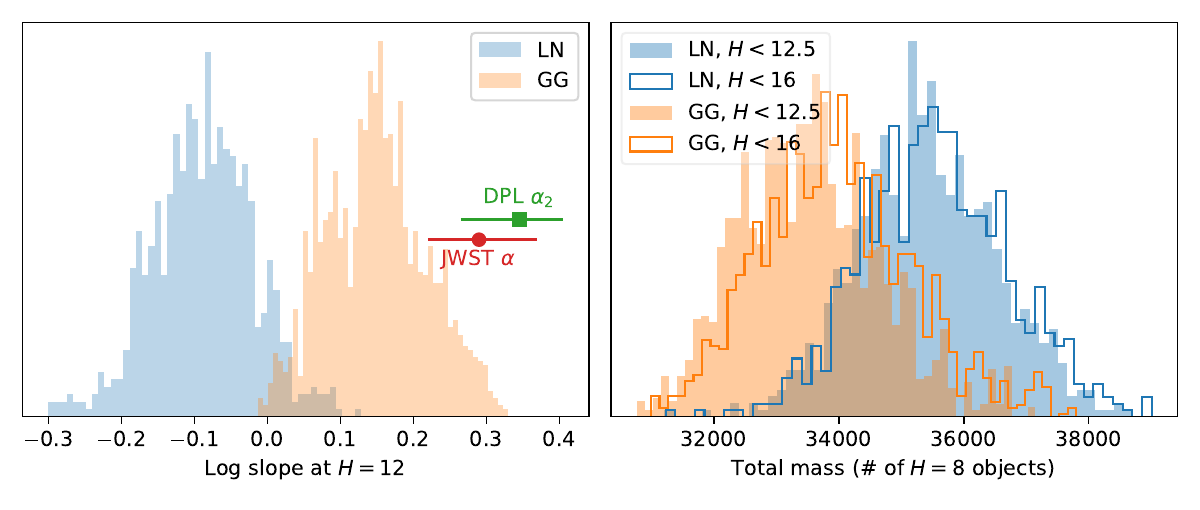}
    \caption{Each panel shows the posterior probability distribution of a property of the CC size distribution under the lognormal (LN) and generalized-$\Gamma$ (GG) models.  At left is the logarithmic slope $d(\log_{10}N)/dH$ at $H_r=12,$, as well as the $\pm1\sigma$ ranges for a double-power-law faint-end slope fit to all the data, and the single-power-law slope fit to only the JWST data \citep{mars}. Both incorporate CC counts several mag brightward of the $11.5<H_r<12.5$ range plotted for the LN and GG models. At right is the integrated mass of the CC's for $H_r<12.5$ (filled) and $H_r<16$ (outlined).  The total mass is in units of the mass of an $H_r=8$ CC, assuming constant albedo and density for the population. The logarithmic slope at the limit of current observations is still model-dependent, but the total mass is determined to $\approx5\%$ independent of the model.}
    \label{fig:dist}
\end{figure}
\subsection{Faint-end behavior}
\label{sec:faint}
The left panel of Figure~\ref{fig:dist} shows the posterior distribution of the local power-law slope of the CC LF near $H_r=12$, under each of the LN and GG models as constrained by the Bright-app collection of observations.  The fact that the two models appear inconsistent with each other is not a contradiction---rather it is a warning to us that \emph{the inferred faint-end slope, in particular whether we have detected a turnover in  the CC LF, is model-dependent.} The DPL/sharp model is an adequate fit to the data with a faint-end slope $\alpha_2\approx0.35,$  above the ranges implied by the LN and GG fits. The model dependence of the faint-end slope is also apparent at the right end of  Figure~\ref{fig:lf}. Since theory offers no compelling constraint on the correct functional form, we can only conclude that the LF flattens considerably by $H_r=12,$ and cannot claim detection of a decreasing $dN/dH.$  

The logarithmic slope of $\alpha=0.29^{+0.08}_{-0.07}$ that \citet{mars} find for the best single-power-law fit to the cold JWST samples lies outside the LN range plotted in blue in Figure~\ref{fig:dist}.  But the two results remain consistent, since we plot the logarithmic slope between $11.5<H_r<12.5,$ while the JWST-only fit is to the \emph{apparent} magnitude distribution $25<m_r<29.5.$ 
The latter extends several magnitudes brighter than the former, so we expect to see a higher mean slope value obtained.

The crater size distribution on outer Solar System bodies such as Pluto, Charon, Triton, and Arrokoth is also a proxy for the size distribution of TNOs, extending to smaller bodies than we have measured directly here.  The $dN/dM$ slopes that we observe near $H_r=12$ here are broadly consistent with the slope derived from the cratering record.  A more thorough comparison will be given in a forthcoming publication.

\section{Mass inferences from $H$ distribution}
\label{sec:H2M}
We summarize here what is known about the average mass $M$ of CCs as a function
of their observed $H_r,$  and what this implies for the total mass of the CC
belt.  For a body near zero phase angle with Earth-facing cross-sectional area $A$ and volume $V$, we
can define a dimensionless shape factor $\shape=3VA^{-3/2}/4\sqrt{\pi},$ which is unity for a
spherical body, $>1$ for a body elongated along the line of sight.
The mass of a body with mean density $\rho$ is then $M=(4\sqrt{\pi}/3)\rho\shape A^{3/2}.$ For geometric albedo $q,$ the absolute magnitude obeys
$10^{-0.4(H-H_0)}=qA$ for some constant $H_0.$   Combining these two yields
\begin{align}
  M & = M_8 \times 10^{-0.6(H_r-8)}, \label{eq:MM} \\
  M_8 & = M_8^\star \shape \times \left( \frac{\rho}{900\,\text{kg}\,\text{m}^{-3}}\right)
        \left( \frac{q}{0.15}\right)^{-3/2} \label{eq:M8} \\
  M_8^\star & =   \frac{4\pi \times900\,\text{kg}\,\text{m}^{-3}}{3} \,
              \left(\frac{1\,\text{au}}{\sqrt{0.15}}\right)^{3}
              10^{-0.6(8-M_{r\odot})} \nonumber \\
              & = 2.31\times10^{17}\, \text{kg} = 3.88\times10^{-8} M_\oplus. \label{eq:M8star} 
\end{align}
where $M_{r\odot}=-26.96$ is the absolute magnitude of the sun in $r$ band \citep{willmer}.  The
$M(H)$ function can thus be described by $M_8,$ which is determined by $\shape,
q,$ and $\rho.$   Note that the geometric albedo $q$ also depends on shape (via
scattering angles and shadowing) for fixed surface properties.
If all CCs were self-similar bodies, \ie\ $q,\shape,\rho$ constant, then $M_8$
would be constant and \eqq{eq:MM} would be all we need.
In reality, differences in shape, albedo, and density lead
to departures from this form, which we can recast as variations in the value of
$M_8.$

\begin{figure}
\includegraphics[width=\textwidth]{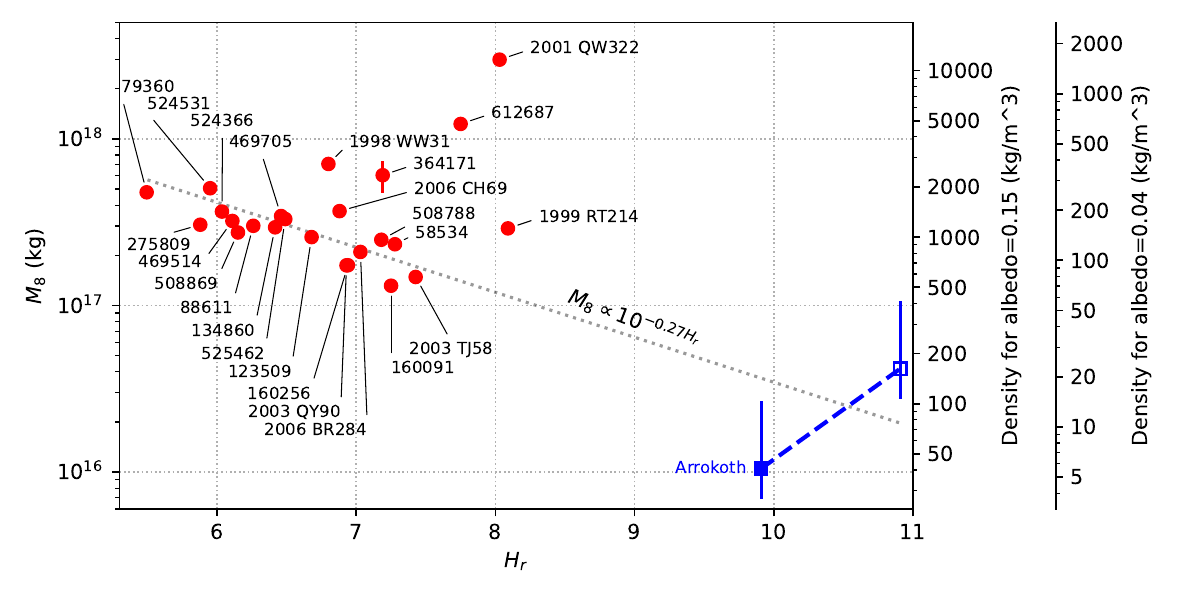}
\caption{The red circles plot the mass normalization factor $M_8$ from
  \eqq{eq:M8} vs the observed absolute magnitude $H_r$ of all CCs with
  well-measured masses, from binary systems. Each is marked with its designation.  $M_8$ is a combination of factors
  related to shape, albedo, and density.  The right-hand axes give the inferred
  bulk density under assumptions of spherical Lambertian bodies with 2 chosen
  geometric albedos.  The solid blue square is the estimated position of
  Arrokoth on this plot; the open blue square is Arrokoth if it were oriented to
  present its minimum brightness toward Earth. The dotted line traces a
  suggestive trend of decreasing $M_8$ with $H.$}
\label{fig:M8}
\end{figure}

\subsection{Known TNO masses vs $H$}
 Figure~\ref{fig:M8} plots the inferred $M_8$ vs $H$ for all CCs with any
 photometry-independent estimate
of mass.  There are zero CCs for which all three of $q,\shape,\rho$ have been
precisely measured.  No CCs have masses determined from spacecraft dynamics, nor from
mutual encounters. All of those in red in Figure~\ref{fig:M8} are binary
systems, as listed in \citet{grundy2015} as updated by W. Grundy (private
communication).  None of these have shape measurements (from stellar
occultations or resolved imaging) and only a few have surface-area estimates from
thermal emission.  Only Arrokoth has a measured shape and accurately inferrable
geometric albedo; but its density (and hence mass) is constrained only by
geophysical considerations of its surface slope \citep{KeaneArrokoth}.
Instead of trying to constrain the value of $M_8$ by
obtaining independent constraints on $q,\shape,\rho$ simultaneously, we will skip these ``middlemen''
and directly consider the $M_8(H)$ relation.

We assume that a binary TNO with a magnitude difference $\Delta H=H_2-H_1$ between
components has the same $M_8$ for both components.  This would be true for matched density, shape, and surface properties---the trend of  matching colors in binaries supports but does not require this condition.  In this case,
 the total flux and mass are
\begin{align}
H_{\rm tot} & = H_1 - 2.5*\log_{10} \left[1 + 10^{-0.4\Delta H}\right], \\
M_{\rm tot} &= M_8 \times 10^{-0.6(H_1-8)} \times \left[1 + 10^{-0.6\Delta H}\right] \\
& = M_8 \times 10^{-0.6(H_{\rm tot}-8)} \times \left[1+10^{-0.4\Delta H}\right]^{-3/2}\times \left[1 + 10^{-0.6\Delta H}\right] \\
\Rightarrow \quad  M_8 & = M_{\rm tot}10^{0.6(H_{\rm tot}-8)} \frac{ \left[1+10^{-0.4\Delta H}\right]^{3/2}}{1 + 10^{-0.6\Delta H}}.
\end{align}
We apply the last factor's adjustment, which is between 1 and $\sqrt{2}$, when estimating $M_8$ from each binary system.  We also recalculate each system's total $H_r$ by averaging the values derived from observations reported to the MPC in $w$, $r$ or $R$ bands (in that order of precedence), following the same procedures as in Section~\ref{sec:bright}.

Arrokoth is the only CC at $H_r>8.5$ with any mass information.
Its mass is estimated using measured surface slopes by
\citet{KeaneArrokoth}.  Its total magnitude in the HST F606W STMAG system as
reported by \citet{benecchiHST} corresponds to $H_r=9.9.$  The blue square in
Figure~\ref{fig:M8} represents Arrokoth at the  density of
$235^{+365}_{-80}$~kg~m$^{-3}$ used by Keane et al.\ to get $1\sigma$
uncertainties on Arrokoth's mass.  The very low value of $M_8$ inferred from
Arrokoth is partly geometrical: this contact binary exposes nearly its maximal
cross-sectional area toward Earth at present (minimal $\shape$), and would appear $>1$~mag fainter
($\shape>1$) if viewed along its major axis \citep{porterShape}, which we plot with the
open blue square.  Its rotation axis is nearly normal to its major axis, though,
so no orientation could ever keep its major axis aligned toward Earth.  Its 
faintest rotation-averaged $H$ value will be $\approx0.7$~mag
fainter than now, between the solid and open blue squares.

\subsection{Values and trends of $M_8$}
The observations present a confusing picture.  One definitive conclusion is that
there are variations of a factor $\approx15$ in the quantity $M_8 \propto \rho
\shape q^{-3/2}$ at fixed $H\approx8,$ and if we take the Arrokoth mass estimate
at face value, there are variations of $\approx100$ in this quantity even if we
factor out Arrokoth's currently ``special'' viewing angle from Earth.  Varying the
geometric albedo from 0.04 to 0.20 changes $M_8$ by $\approx 11.$  The bulk
density might vary from 200--1200~kg~m$^{-3},$ changing $M_8$ by factor $\approx6.$  The
factor $S$ (essentially the ratio of line-of-sight axis to the geometric mean of
the other two) is plausibly $0.5<\shape<2,$ a factor of $4.$  Spanning the $>100\times$ range of
$M_8$ between Arrokoth and 2001~QW$_{322}$ cannot be done by altering only one of
these three factors to its extremes---indeed it requires pushing nearly all
three of these factors to their opposite plausible limits in these two bodies,
or revising our opinion of plausible CC densities and albedos.
At least two  
of the binaries require unreasonably high densities if their albedos are
$\approx0.15$ and they are nearly spherical.  For 2001~QW$_{322}$, the density is
uncomfortably high even with albedo 0.04, further suggesting that the component
bodies of this system must be very non-spherical, and/or have an unusual
history, such as being a pair of fragments of a differentiated body.  An
irregular shape with the longest axis along the line of sight ($\shape>1$) can raise a body's location on this plot, but maintaining this orientation would also require the spin axis to be along the line of sight.  For a contact binary, we expect the spin axis to be normal to the longest axis, as for Arrokoth.

The mean value of $M_8$ of the binaries is $(2.0\pm0.5)\,M_8^\star,$ taking the
uncertainty from the observed scatter.  This formal uncertainty is an
underestimate, given the very non-Gaussian
distribution of $M_8.$ The median value is only $1.3M_8^\star.$ Also there is
the possibility of strong selection biases on $M(H)$ for CCs that are detectable
binaries, and have successful followup.

All but 5 of the 24 binaries lie in a fairly tight locus showing a strong
downward trend in $M_8$ vs $H_r.$  It is intriguing that the extrapolation of
this $M_8 \propto 10^{-0.27 H}$ trend roughly matches Arrokoth at its more
typicial viewing angles.  Some trend of this nature could result from
gravitational compaction of larger bodies, but it is premature to conclude that
this process is occurring, or that it would extend down to bodies as small
as $H_r>8$.  The outliers complicate any conclusion about
the universality of the trend.  In fact for the binaries, the mean $M_8$ is
actually rising to fainter $H$ if we include the outliers.
Further investigation of 2001~QW$_{322}$ and (612687) via spectroscopy, radiometry, and light-curve analyses could be very illuminating.

If this trend is a real characteristic of the full population, it amounts to a
change in $M_8$ of a factor $\approx5$ in the 2.5-mag range of $H$ spanned by
the binaries, and a factor $\approx20$ if it continues to Arrokoth.  These
values strain and exceed, respectively, the amount of $M_8$ change we might
attribute to compaction alone.  Shape and albedo trends would also need to be
present.   Further investigation is clearly warranted, but we defer it to a
future publication.

This trend in $M_8(H)$ would also imply that the 8~mag span $5\lesssim H_r \lesssim 13$
over which we have CC detections corresponds to 7 orders of magnitude spread in
mass, instead of the 4.8 orders of magnitude for constant $M_8$.

\subsection{Total CC mass}
\label{sec:mass}
The right panel of Figure~\ref{fig:dist} plots the posterior distribution of the
integrated mass of the CCs, under the assumption that $M\propto 10^{-0.6H_r},$
\ie\ $M_8$ is constant.
This is seen to be well constrained under either the LN or GG models, which now agree within uncertainties.  The resulting $M_{\rm CC}=(3.5\pm0.2)\times 10^4 M_8$ is robust between these models, and both models show that $>99\%$ of the mass is at $H<12.5$.  The observational data do not preclude a strong upturn in the LF at $H>12.5,$ \ie\ a substantial reservoir of mass in bodies of diameter $d\lesssim 10$~km. 

If we ignore any possible km-scale mass reservoir, and assume spherical bodies ($\shape=1,$), then the inferred total mass of the CCs is
\begin{equation}
    M_{\rm CC} = (1.35\pm0.07)\times10^{-3}
    M_{\oplus} \left(\frac{q}{0.15}\right)^{-3/2} \left(\frac{\rho}{900\,\rm{kg}\,\rm{m}^{-3}}\right).
\label{eq:mCC}
\end{equation}
This agrees with and sharpens the estimate of the prefactor of $1.5^{+0.9}_{-0.4}$ for $H_r<12$ from \citet{deepV}, and $1.1\pm0.2$ from \citet{bernstein2004size}.

The uncertainty in CC mass is clearly dominated by uncertainty in the $M_8(H)$
relationship, \ie\ $\rho\shape q^{-3/2}.$   Above we found median (mean) values of
$M_8=1.3\,(2.0)\,M_8^\star,$ implying total CC masses of
1.7--2.7$\times10^{-3}M_\odot.$ If there is a downward trend in $M_8(H)$, the
integrated
mass would be below these values.

\section{Discussion}

\subsection{Comparison to previous measurements}
\label{sec:lfmeas}
Recent estimates of the $H$ distribution of CCs include:
\begin{itemize}
\item \citet{fraser2014absolute} fit a sharp DPL [\eqq{eq:dpl} with $\beta\rightarrow\infty$] with parameters $(\alpha_1, \alpha_2,H_b)=(1.5,0.38,6.9),$ to observations including the CFEPS and HST observations.
\item \citet{JJ} fit the TCP model of \eqq{eq:ggcdf} with preselected values of $\alpha=0.4,0.5$ to the OSSOS++ detections at $5\lesssim H_r \lesssim 8.3$ and note consistency of the extrapolation to the bright (MPC) and faint (HST) data. The fitted parameters are $(\beta,H_b)=(0.25,8.1)$ and $(0.35,7.1)$ for the 2  choices of $\alpha.$
\item \citet{deepV} find that LN, TCP, sharp DPL models, and even single power laws can all satisfactorily fit the preliminary discoveries from the DEEP survey at $6 \lesssim H_r \lesssim 10.5.$  We will compare to their LN fit, which is equivalent to \eqq{eq:roll} with $(\alpha_1, \alpha_2)=(0.77,-0.10).$
\item \citet{des2} fit TNO subsets discovered by DES in the range $5.5<H_r<8.2,$ agreeing with the conclusion of  \citet{petitLF} that TNOs in this range have the same LF shape across the hot and cold classical populations.  The most precise LF constraint from DES fits the full population to an LN form with $(\alpha_1, \alpha_2)=(0.39\pm0.03,-0.23\pm0.05).$
\end{itemize}
\begin{figure}
\centering
\includegraphics[width=0.8\textwidth]{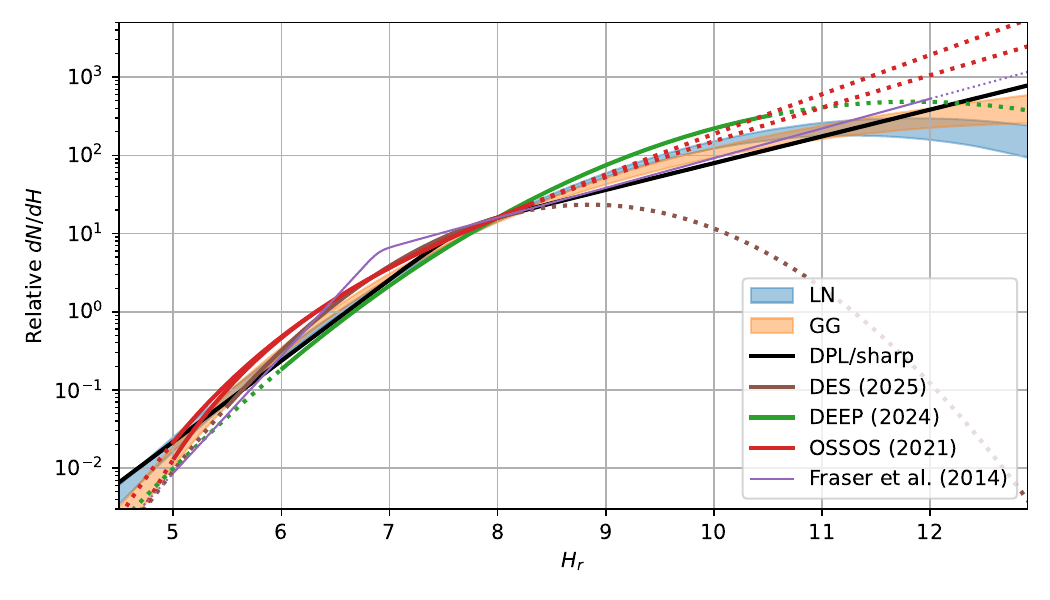}
\caption{Fits to the $dN/dH$ distribution of CC's from different data collections.  The LN (blue) and GG (orange) bands are our 68\% credible regions under each model.  Other curves have been normalized to match these at $H_r=8.$ The  DPL/sharp curve 
is our maximum-likelihood abrupt DPL fit,  giving an idea of the variation allowed from choice of analytic form.  The other curves are best-fit values from previous works enumerated in Section~\ref{sec:lfmeas}.  The solid portions of each curve are the range of the data used for each fit; the dotted portions are extrapolations. There are two red curves, plotting the the OSSOS++ fits at their two preselected faint-end slopes \citep{JJ}.}
\label{fig:lfmeas}
\end{figure}

Figure~\ref{fig:lfmeas} plots these previously derived CC LFs along with our allowed ranges for the LN and GG models, plus the best-fit sharp-transition DPL. The previous results generally fall as close to our range as expected from the $H$ range and size of the CC samples that they fit. The JWST measurements allow significant improvement over the DEEP and CFEPS+HST results for $H_r\gtrsim 8.$

Our derived LN parameters are just outside the $2\sigma$ contours for the DES fit to $5.5<H_r<8.2$ by \citet{des2}, toward a slower roll.  As a consequence our CC total mass estimate is higher than their extrapolation of the total CC mass in $5<H_r<12$ of $0.67^{+0.10}_{-0.08}\times10^{-3}M_\oplus$ under the same albedo and density assumptions as we make in Section~\ref{sec:mass}.

We defer to a future publication the detailed comparison of the observed CC distribution to the mass distribution of impactors derived from crater size distributions on outer-solar-system bodies seen in resolved spacecraft imaging.

\subsection{Comparison to streaming instability simulations}
\label{sec:sims}
Detailed comparison of the observed CC LF to theoretical predictions for SI is not yet productive because the theory is not strongly predictive.  The analytical expression for the expected $dN/dM$ from \citet{GerbigLi} assumes that this function is proportional to the growth rate of perturbations after initial concentration via SI or other mechanisms. As the authors note, further investigation is needed to understand the simultaneous action of the concentration mechanism and the gravitational collapse phase.

Other theoretical predictions are functional forms that are fit to masses of pebble concentrations in hydrodynamic simulations.  The simulations are extremely numerically intensive, and \citet{Schafer2024} have shown that the resultant LFs are substantially dependent on the volume and resolution of the simulation in the ranges typically used, even with the initial physical state fixed. 

Even with converged simulation results, there are additional processes not fully included in the simulations that could strongly influence the observed LF, such as binarity and other forms of multiplicity/loss during gravitational collapse of the clouds \citep{robinson}. There is additional dependence on the algorithms used to define the pebble clumps.

\begin{figure}
\centering
\includegraphics[width=0.6\textwidth]{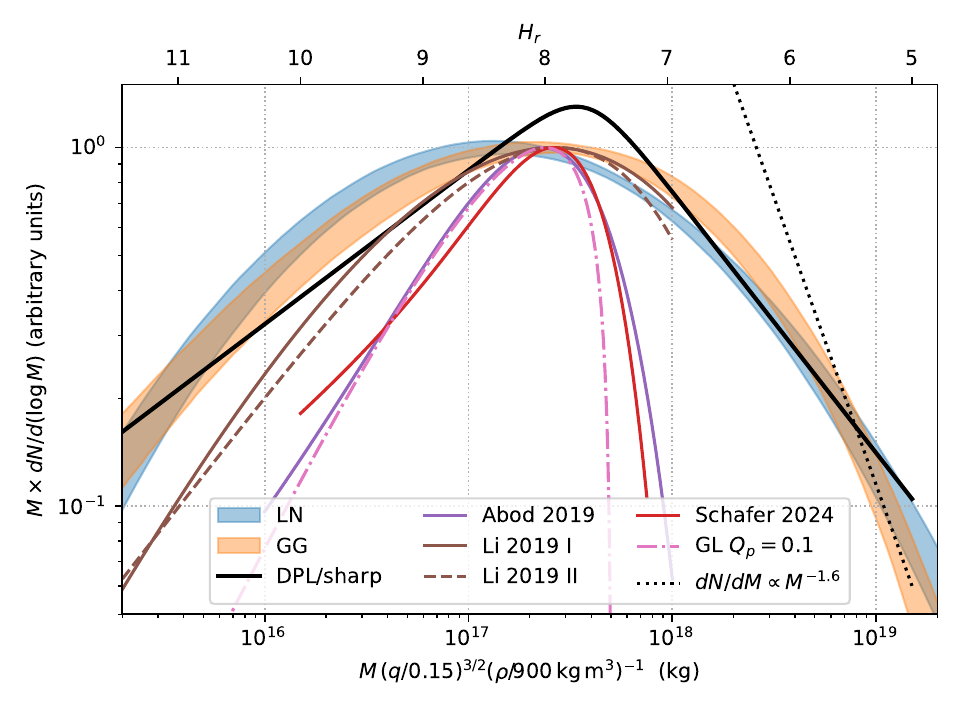}
\caption{The quantity $M\,dN/d(\log M)$ is plotted vs $H_r$ and $M$ for the LN, GG, and sharp-cornered DPL models fit to the observations.  The $y$ axis plots the mass per logarithmic interval in mass, in arbitrary units. The $x$ axis is CC mass assuming nominal albedo $q$ and density $\rho.$ We overplot several LFs fit to results of numerical SI instabilities, as detailed in the text.  Each curve spans the range of mass of pebble clouds found in the simulation, but with arbitrary shifts in the $x$ and $y$ axes to facilitate comparison to the true CC shape.}
\label{fig:masssims}
\end{figure}

Instead of detailed numerical comparison of the CC LF to theory, we offer a visual comparison in Figure~\ref{fig:masssims} of the observed CC mass distribution to those derived from the more recent SI simulations/theory. The plotted quantity $M\, dN/d(\log M)$ gives the population mass per log interval in body mass, \ie\ how broadly the total CC mass is distributed among relative masses.  Uncertainty ranges for the LN and GG fits to the data are shown---a LN distribution is a parabola in this plot. The best-fitting sharp DPL (\ie\ broken power law) is also plotted---power laws are straight lines.  The other curves plot the $M\,dN/d(\log M)$ of published fits to SI simulations or theory.  Each of the following is plotted, with an arbitrary shift on both axes to enable visual comparison of the shapes.  \citet{liChiang} note that most simulations to date do not yield total CC masses comparable to the real CC, and therefore substantial shifts in initial conditions---and hence in the typical masses and belt densities---will be necessary to match reality.
\begin{itemize}
\item The only simulation with number of pebble clouds substantially larger than the number of CCs in our sample is that of \citet{Schafer2024}.  We plot their favored parameters for the TCP model fit to their highest-resolution $L=1.6H$ model with the ``improved'' sink particle creation algorithm: $N(<M) \propto M^{-\alpha} e^{-(M/M_b)^\beta}$ with $(\alpha, \beta)=(0.51,1.48).$
\item \citet{abod2019mass} fit the TCP model with fixed $\beta=1$ and find $\alpha\approx0.3$ fits a range of their simulations.
\item \citet{li_demographics_2019} demonstrate that a variety of analytic forms can fit their SI results nearly equally well.  If we exclude their models that imply discontinuous $dN/dM$ functions, then the TCP fits are as good as or better than other functional forms, so we plot these for both their runs I and II.
\item We overplot the simple power law distribution of mass that was fit to some earlier simulations \citep{simon_mass_2016}  It is not consistent with the CCs.
\item The dash-dotted curve in Figure~\ref{fig:masssims} plots the analytic theory of \citet{GerbigLi} for a dimensionless parameter $Q_p=0.1,$ scaled in mass and number to match the peak to the data.  Values such as this near zero yield the broadest $dN/d\log M$ functions, whereas $Q_p\rightarrow1$ narrows the predicted function.  As its original authors note, the analytic solution resembles many simulation results.
\end{itemize}
Each simulation is plotted only over the span of masses that were produced by the simulation.  We note that the 5 orders of magnitude of CC mass now measured observationally exceeds the dynamic range achieved in the simulations, which is not surprising given that dynamic range increases simulation cost.

The immediate conclusion is that \emph{all of the SI simulations [except for those of \citet{li_demographics_2019}] have bright-end declines that are too steep to match the CC population. Additionally, all have mass distributions that are too narrow in $\log M,$  (with Li et al. again closer to the observations).}  This is also evident in the lower values of $\beta$ we derive for the TCP model in \eqq{eq:pTCP}---even considering the substantial degeneracies evident in Figure~\ref{fig:ggmh}---than are reported in fits to SI simulation cloud masses.

The simulations' results could be broadened to better match the data if we were to assume spatial or temporal variations across the CC birth zone/era of the parameters that drive the mass scale.  A more fine-tuned form of birth inhomogeneities would be needed to turn the steep simulated high-mass slopes into the shallower observed distributions.  

Section~\ref{sec:H2M} shows data suggesting that the $M(H_r)$ function is
 in fact steeper than $10^{-0.6H_r},$ \eg\ as would occur from gravitational
 compaction of porous pebble piles.  As noted there, this would imply that the
 $M dN/dM$ function would be even broader than plotted here,, increasing the discrepancy between SI simulations and the true CC differential mass distribution.

 After this paper was posted and submitted, \citet{Hall2026} presented detailed comparison between the observed CC LF (including the data herein) and clump mass distributions from a new suite of SI simulations.  After allowing for mass and number scaling, one simulation does have a K-S statistic allowing for the observations and the simulations to share a common distribution, with potentially better agreement at the high-mass end ($\beta$ value of the TCP form) than we see in Figure~\ref{fig:masssims}.

\subsection{Near-future observations}
The deviations between the best-fit LN, GG, and DPL models for $dN/dH$ are 20--30\% in the $H_r$ range where forthcoming analysis of the DEEP observations should strengthen constraints substantially. The completed DEEP survey will surpass the
 preliminary results of 
\citet{deepV} by having full orbital solutions fit to multi-epoch detections of $>1000$ CCs extending $\approx 2$~mag fainter than OSSOS. Forthcoming analysis of the now-complete CLASSY survey \citep{classy} will likewise greatly expand the number of known CCs beyond limits of DES and OSSOS.

Over the next 2--3~years, the Legacy Survey of Space and Time will detect and characterize many thousands of objects \citep{kurlander} covering essentially the entire sky footprint of the CCs, removing any uncertainty about the behavior of the population for $H_r\lesssim9.$  This would enable detection of subtle structure in the LF, if for example there are ``waves'' as seen in the MBA size distribution that are attributed to collisional grinding \citep[see review by][]{Bottke2015}. The potential presence of LF waves in TNO-derived populations such as the Jupiter Trojans is discussed by \citet{bottke2023}.

Additional observations with the JWST NIRCAM (Program  GO-7700) will  extend $\approx2$~mag deeper than the \citet{mars} data, or a factor $\approx15$ smaller mass.  These data should detect any substantial upturn in $dN/dH$ for $\approx5$~km bodies, \eg\ small fragments ejected from collapsing pebble clouds \citep{robinson}.

\section{Conclusions}
\label{sec:conclusions}
The observations in Table~\ref{tab:surveys} can be fit by a variety of functional forms, the simplest of which is \eqq{eq:roll} which implies a lognormal distribution of CCs' masses.  Even across the factor $10^5$ of mass that the data span, multiple analytic forms for $dN/dM$ give valid fits (Figure~\ref{fig:lf}), but with more accuracy in the parameters than in previous analyses (Figure~\ref{fig:lfmeas}).  Within the range of the data, these functions agree to $\lesssim30\%$, but diverge substantially at $H_r>12.5.$  The analytic fits such as the lognormal should be viewed as descriptive of nature within the range of observations, not predictive or definitive. Extrapolation of the functional forms to $H_r\gtrsim 12$ is extremely model-dependent and ill-advised.

For this section we will assume that CCs are spherical bodies with albedo $q=0.15$ and density $\rho=900$~kg~m$^{-3},$ which Section~\ref{sec:H2M} demonstrates yields a plausible match to the observed $M$ vs $H_r$ relation for CC binaries. In this case our current detection limits of $H_r\approx13$ correspond to diameter of $D\approx 8$~km and mass $M\approx2\times10^{14}$~kg.  
Barring a strong upturn in $dN/dH$ or $dN/dM$ for objects smaller than
$H_r\approx13,$ we know the \emph{number} of CCs vs $H$ very well, and the
uncertainties on the inferred total mass of the CCs is dominated by our
ignorance of their mass vs $H$ relation---we are even more ignorant of the
individual albedo, shape, and mean density factors that determine $M$ vs $H.$
The data in Section~\ref{sec:H2M} lead to an estimate of total CC mass
1.7--2.7$\times10^{-3}M_\odot.$  The quantity $M\, dN/d(\log M),$
which gives the total CC mass per logarithmic interval of body mass, peaks near
$2\times10^{17}$~kg (Figure~\ref{fig:masssims}), placing this typical CC mass
near $H_r=8, D=75$~km.  The largest known CC,\footnote{The LSST will determine
  whether there are any larger CCs in the next 1--2 years.} 2014~TD$_{86}$ at
$H_r=5,$ is $\approx60\times$ more massive than the typical mass---and moreso if
it has been compacted to higher density than typical.   

Theories of planetesimal formation will need to be capable of reproducing the observed CC $dN/dM.$ Because of the 
complex multiscale physics of the formation process, analytic predictions for the mass distribution given initial conditions are limited, and numerical simulations are limited in volume and dynamic range.  Nonetheless we can compare the CC mass distribution to state-of-the-art simulations of formation of planetesimals via the streaming instability (SI) followed by pebble-cloud collapse (Figure~\ref{fig:masssims}).  The simulations and the analytic formulation from \citet{GerbigLi} generally find much smaller ratios between the typical mass and the maximum mass, \ie\ steeper distributions at the high-mass end.  Some of this is plausibly due to the small simulation volumes (which suppresses the maximum mass) and dynamic-range limitations.  But the bright-end slope is still larger in all SI simulations---except those of \citet{li_demographics_2019}---than in the CCs.

The CCs have mass distributed over a broader range of $\log M$ than indicated by most or all of the SI simulations.  This does not by any means invalidate the SI hypothesis.  The simulation results still vary substantially with choices of initial conditions and algorithms,  do not incorporate some possible or likely physical processes such as fragmentation of pebble clouds upon collapse \citep{nesvorny2021, robinson}, and also have substantial uncertainties from counting statistics.  In addition, the mass distribution can readily be broadened by introducing inhomogeneity of the initial conditions of the CC gas/dust disk.  

The CC mass distribution inferred herein can thus be considered an empirical target for future simulations to match.  Near-future observations will allow extension of the measured $dN/dH$ to smaller bodies, and greatly lower the statistical uncertainties for CC sizes amenable to ground-based detection.  Either could detect revelatory features not well described by the simple lognormal model that fits current data.

%% Please use the acknowledgment and contribution environments. This will 
%% be anonomyized when the "anonymous" style option is used. 
\begin{acknowledgments}
We thank J.J. Kavelaars and Jean-Marc Petit for providing detailed information on the OSSOS detections and on the choices made in \citet{JJ}.

This work is based on observations made with the NASA/ESA \textit{Hubble Space Telescope} and the NASA/ESA/CSA \textit{James Webb Space Telescope}. The data were obtained from the Mikulski Archive for Space Telescopes at the Space Telescope Science Institute (STScI), which is operated by the Association of Universities for Research in Astronomy, Inc., under NASA contracts NAS 5-26555 (HST) and NAS 5-03127 (JWST). 

Support for US authors of this paper  was provided by NASA via STScI grants to programs HST-GO-16720 and JWST-GO-01568.

G.M.B. additionally acknowledges support from the National Science Foundation under grant AST-2407527.

W.C.F. and M.R.E. acknowledge funding from the CSA via grants 222JWGO1-09 and 22EXPCO11. This work also used the facilities of the Canadian Astronomy Data Centre (CADC), operated by the National Research Council of Canada (NRC) with support from the CSA.

M.J.H. and K.J.N. acknowledge funding from NSF grant AST-2206194 and NASA Yearly Opportunities for Research in Planetary Defense (YORPD) Program grant 80NSSC22K02.

This research has made use of data and/or services provided by the International Astronomical Union's Minor Planet Center. 

\end{acknowledgments}

\begin{contribution}
GB led the likelihood fitting and drafting of the paper, with WF and KN also contributing substantially.  WF contributed to gathering the archival data, and he and AM determined appropriate color transformations between surveys. WG contributed the information on masses of CCs.  All authors contributed to the planning, execution, and analysis of the JWST ``Pencil Beam'' TNO survey that made this analysis possible, and discussions of the interpretation and presentation of the results.

\end{contribution}

\facilities{JWST(NIRCam)}
\software{astropy \citep{2013A&A...558A..33A,2018AJ....156..123A,2022ApJ...935..167A}, scipy \citep{scipy}, numpy \citep{numpy}}

\bibliography{cclf}{}
\bibliographystyle{aasjournalv7}

\end{document}